\documentclass[lettersize,journal]{IEEEtran}
\usepackage{amsmath,amsfonts}
\usepackage{amsthm}
\usepackage{algorithm}
\usepackage{algpseudocode}
\usepackage{array}
\usepackage{enumitem}
\usepackage[caption=false,font=normalsize,labelfont=sf,textfont=sf]{subfig}
\usepackage{textcomp}
\usepackage{stfloats}
\usepackage{url}
\usepackage{verbatim}
\usepackage{graphicx}
\usepackage{multirow}
\usepackage{booktabs}
\usepackage{tabularx}
\usepackage{cite}
\newtheorem{lemma}{Lemma}[section]
\newtheorem{corollary}[lemma]{Corollary}

\begin{document}

\title{Constraint-based Adversarial Example Synthesis}

\author{
\begin{tabular}{@{  }c@{  }c@{  }c@{  }}
  \IEEEauthorblockN{Fang Yu\textsuperscript{1}} & \IEEEauthorblockN{Ya-Yu Chi\textsuperscript{2}} & \IEEEauthorblockN{Yu-Fang Chen\textsuperscript{3}}\\
\IEEEauthorblockA{
\textit{Department of Management} & \textit{Department of Management} &\textit{Institute of Information}\\
\textit{Information Systems} & \textit{Information Systems} &\textit{Science}\\
\textit{National Chengchi University} & \textit{National Chengchi University} & \textit{Academia Sinica}\\
Taipei, Taiwan & Taipei, Taiwan & Taipei, Taiwan\\
\textsuperscript{1}yuf@nccu.edu.tw & 110356043@g.nccu.edu.tw\textsuperscript{2} & \textsuperscript{3}yfc@iis.sinica.edu.tw}
\end{tabular}




}



\maketitle

\begin{abstract}
In the era of rapid advancements in artificial intelligence (AI), neural network models have achieved notable breakthroughs. However, concerns arise regarding their vulnerability to adversarial attacks. This study focuses on enhancing Concolic Testing, a specialized technique for testing Python programs implementing neural networks. The extended tool, PyCT, now accommodates a broader range of neural network operations, including floating-point and activation function computations. By systematically generating prediction path constraints, the research facilitates the identification of potential adversarial examples. Demonstrating effectiveness across various neural network architectures, the study highlights the vulnerability of Python-based neural network models to adversarial attacks. This research contributes to securing AI-powered applications by emphasizing the need for robust testing methodologies to detect and mitigate potential adversarial threats. It underscores the importance of rigorous testing techniques in fortifying neural network models for reliable applications in Python.
\end{abstract}


\section{Introduction}
In recent years, deep neural networks (DNNs)~\cite{b1} have achieved notable success in diverse domains such as computer vision~\cite{b2}, speech recognition~\cite{b3}, and game playing~\cite{b4}. While their widespread adoption in safety-critical and socially sensitive applications underscores their significance, it also raises concerns about software verification and security. Despite their generalization capabilities, DNNs often exhibit unexpected behaviors when exposed to carefully crafted perturbations~\cite{b5,b6,b7}, posing a significant challenge to ensuring the dependability of neural network systems.

This paper focuses on the synthesis of constraint-based adversarial examples, addressing the challenges associated with the verification and security of deep neural networks. Previous research has explored formal verification techniques, utilizing symbolic representation and formalization of neural network computation. Approaches like Reluplex~\cite{b8} and Marabou~\cite{b9} leverage SMT solvers for verifying fully connected and convolutional neural networks, but scalability limitations necessitate exploration of alternative methods, including automatic testing.

Automatic testing techniques, exemplified by DeepXplore~\cite{b10}, DeepGauge~\cite{b11}, and DeepMutation~\cite{b12}, offer an alternative avenue by providing evidence to support safety arguments. These methods aim to generate adversarial examples that expose vulnerabilities in neural networks. Concolic testing, guided by coverage metrics like neuron coverage, has been employed to systematically explore and increase the coverage of neural network behaviors~\cite{b13,b14,b15,b16}. However, challenges arise in terms of scalability and the effectiveness of coverage metrics.

To address these challenges, we extend the Python concolic tester PyCT~\cite{b17} to handle various neural network components, including floating-point computation, connected layer computation, convolution, recurrent neural networks, long short-term memory, max pooling, softmax, and non-convex activation functions (e.g., ReLU, tanh, Sigmoid). Our extended PyCT incorporates symbolic execution on specific input positions and branches, mitigating the computational burden of constraint solvers. Leveraging DeepSHAP~\cite{b18}, we estimate the influence of input features, enabling PyCT to focus on inputs contributing most to network predictions. This targeted exploration of influential inputs enhances the efficiency of concolic testing for adversarial example generation.

In summary, our work advances constraint-based adversarial example synthesis by extending PyCT's capabilities and incorporating DeepSHAP. Through empirical evidence, we demonstrate the effectiveness of our approach in reducing computational costs and enhancing the scalability of adversarial example generation.

In conclusion, our extended PyCT~\cite{b19} for synthesizing constraint-based adversarial examples in deep neural networks aims to contribute comprehensive insights and data to advance research in the detection and defense against adversarial attacks. We anticipate that our methodology will inform future endeavors, fostering a more secure landscape and facilitating the development of robust strategies for detecting and mitigating adversarial threats in neural networks.

\section{Related Work}
\subsection{Concolic Testing}
Sen et al.~\cite{b20} define Concolic Testing as a method that combines concrete and symbolic execution of the code under test to automate test input generation. Cha, S. et al.~\cite{b21} further elaborate on Concolic Testing, introducing a symbolic memory state $S$ and a path condition $\Phi$. Here, $S$ functions as a mapping for variables and symbolic values, such as the association of variable $a$ with $\beta$. Concurrently, $\Phi$ represents the path condition, expressed as $\Phi = \phi_1 \wedge \phi_2 \wedge \phi_3 \dots$ for the current test input. The process begins with concrete execution using a random input, generating constraints $\Phi = \bigwedge_{j\leq i} \phi_j$ on the program's input. Subsequently, the last conjuncted constraint is negated, enabling symbolic execution to generate a new input with fresh path constraints $\Phi^{'} = \bigwedge_{j<i} \phi_j \wedge \neg \phi_i$, potentially resulting in different program outputs.

Concolic testing, through the synergy of concrete and symbolic execution, offers comprehensive test coverage, automates test case generation, and exposes potential errors, including boundary conditions and security vulnerabilities. Its applicability spans diverse domains, including software security, smart contract testing, and embedded systems. Kim et al. present FOCAL~\cite{b22}, an innovative concolic testing technique that efficiently detects bugs by combining unit and system-level testing. FOCAL uses concolic unit testing to identify target failures and generates system-level test inputs for validation. Employing a target-driven refinement technique, FOCAL detects 71 out of 100 bugs and 13 new crashes in empirical evaluations, supported by real-world bug data for SIR benchmarks. It achieves a balance between quick bug identification and validation within a limited testing timeframe. Fortz et al.~\cite{b23} leverage concolic testing to generate test inputs for logic programming languages, addressing the scarcity of robust concolic testing tools in this domain. They propose an enhanced scheme supporting negative constraints and formulate selective unification problems as constraints on Herbrand terms. Applied to Prolog, this approach utilizes the SMT solver Z3 for efficient and scalable solutions to selective unification problems, enhancing the completeness and speed of test input generation. Meng et al. introduce RTL-ConTest~\cite{b24}, a concolic testing framework that efficiently detects security vulnerabilities in System-on-Chips (SoCs). Operating directly on RTL, it circumvents state space challenges and generates security-specific test cases using a novel CFG generator. Evaluated on RISC-V-based SoCs, it successfully identifies various vulnerabilities, offering a scalable and direct solution for securing embedded systems.

Concolic testing excels in managing complex systems, enhancing testing efficiency, and fortifying code reliability through symbolic execution. Its automated methodology accelerates the testing pipeline, proving to be an invaluable asset for swift issue detection and effective resolution in software development. The comprehensive coverage it provides ensures a thorough exploration of diverse execution paths, uncovering hidden errors and vulnerabilities. In essence, concolic testing stands as a versatile and powerful technique for advancing software quality and mitigating potential risks.

\subsection{Coverage-guided Testing on Neural Network}
Coverage-guided testing, a traditional practice in assessing and improving software application quality, involves guiding the generation of test inputs based on coverage achieved during program execution. However, with the increasing influence of deep neural networks (DNNs) across diverse domains, the application of coverage-guided testing to neural networks has emerged as a focal point of active research and development.

In 2018, Sun et al.~\cite{b25} pioneered the application of Concolic Testing to Deep Neural Networks, introducing the "neuron coverage" metric to gauge the percentage of activated neurons in a network. This metric, demonstrated to be a robust indicator of test effectiveness, guided the development of a concolic testing tool tailored for DNNs called DeepConcolic~\cite{b14} in 2019. DeepConcolic, a symbolic reasoning-based tool, explores decision boundaries and crafts adversarial inputs to expose vulnerabilities in DNNs. By combining gradient-based and constraint-solving methods, DeepConcolic systematically maximizes neuron coverage across various paths, contributing significantly to the realm of constraint-based testing for neural networks.

While DeepConcolic focuses on inactive neurons, Zhou et al. introduced a novel coverage criterion, DeepCon~\cite{b26}, and an associated test generation tool, DeepCon-Gen, designed specifically for DNNs. In contrast to DeepConcolic, which targets inactivated neurons, DeepCon introduces contribution coverage—a criterion that integrates neuron outputs and connection weights, providing a more comprehensive understanding of DNN behavior.

DeepCon-Gen prioritizes guided test generation by employing the gradient ascent optimization algorithm, transforming the test generation problem into an optimization challenge. The primary objective is to activate inactive contributions, utilizing constraint and penalty strategies to ensure simultaneous activation of both contributions and neuron outputs. This approach effectively enhances testing adequacy by guiding the test generation process through contribution coverage. Considering both neuron activation and connection weights, DeepCon-Gen facilitates a nuanced exploration of the DNN decision space, contributing to a more thorough assessment of DNN behavior. 

Compared to previous work, we propose leveraging concolic testing and deepshap techniques to systematically explore pivotal branch decisions influencing network predictions along the inference execution and generate crucial test inputs on critical pixels for adversarial example creation.

In addition to constraint-based testing, fuzzing testing is also commonly applied to DNNs. DeepHunter~\cite{b27} introduces a comprehensive fuzz testing framework for evaluating the testing coverage of deep neural network (DNN) models, incorporating five distinct coverage criteria: Neuron Coverage (NC), k-Multisection Neuron Coverage (KMNC), Neuron Boundary Coverage (NBC), Tight Neuron Boundary Coverage (TNBC), and k-Multisection Neuron Boundary Coverage (TKNC). These criteria provide a nuanced assessment of potential defects and vulnerabilities within DNNs, enhancing testing coverage.

In the context of adversarial fuzz testing, DeepHunter employs a morphological mutation strategy to create meaningful variations while preserving the semantic meaning of the original input. This fuzzing strategy involves selecting initial test input seeds and applying controlled mutations to generate new test inputs, adhering to morphological mutation constraints that ensure validity and reasonableness. By challenging the robustness of DNN models with adversarial test cases, DeepHunter provides valuable insights into potential vulnerabilities while avoiding the generation of unrealistic or invalid inputs. To achieve increased mutation changeability during the fuzzing process while generating semantics-preserved tests, DeepHunter selects from two categories of transformations: Pixel Value transformation (changes pixel values) and Affine transformation (moves pixels of the image). The metamorphic mutation strategy balances mutation changeability and semantics preservation through specific definitions (e.g., L0 and L$\infty$) and constraints, ensuring generated tests retain validity and preserve semantics. The metamorphic mutation algorithm in DeepHunter takes an image s and the total number of images to be generated (K) as input, producing a set of newly generated tests (T) as output. This approach ensures controlled and meaningful transformations during the adversarial fuzz testing process.

However, existing metrics are primarily designed for convolutional neural networks (CNNs), rendering them insufficient for recurrent neural networks (RNNs) due to their neglect of internal structures and, more critically, the temporal relations inherent in RNNs.

To address these limitations, Huang et al. propose dedicated structural coverage metrics specifically tailored for Long Short-Term Memory networks (LSTMs), a crucial subclass of RNNs, along with the integration of fuzz testing elements through the introduction of a test case generation tool called TestRNN~\cite{b13}. The introduced metrics, including Temporal Coverage (TC), Boundary Coverage (BC), and Stepwise Coverage (SC), offer a unique perspective by quantifying multistep temporal relations, providing insights into the dynamic internal behavior of LSTM cells processing sequential inputs. Implemented in a prototype tool named TestRNN, these metrics utilize two distinct test case generation algorithms: random mutation and genetic algorithm-based targeted mutation. The targeted mutation process strategically leverages coverage knowledge to guide test case generation, fostering diversity initially and shifting to targeted mutation when coverage improvement plateaus. This process generates corner test cases aimed at enhancing coverage in specific critical areas, crucial for uncovering potential faults. Rigorous evaluation through extensive experiments on various LSTM benchmarks demonstrates the efficacy of the proposed metrics and the TestRNN tool in advancing the understanding and testing of LSTMs.

\subsection{Adversarial Attack}
Adversarial attacks on neural networks have become a critical concern in machine learning, notably emphasized by the seminal work of~\cite{b28}. This concept involves introducing subtle perturbations to inputs, leading the model to confidently misclassify them. Szegedy et al. demonstrated that neural networks often learn input-output mappings with discontinuities, making them susceptible to adversarial attacks. Notably, the Fast Gradient Sign Method (FGSM)~\cite{b29} exploits the linearity in high-dimensional spaces, revealing the efficacy of leveraging this linearity for adversarial attacks in image recognition tasks.

Moosavi-Dezfooli et al. proposed DeepFool~\cite{b30}, which iteratively determines minimal changes to induce misclassification by adjusting the input. The Universal Adversarial Perturbation (UAP)~\cite{b31} is introduced as a technique that aggregates small perturbations across a set of training points, aiming to discover a universal perturbation that induces misclassifications in natural images. These perturbations are imperceptible to humans but impactful for machine learning models.

As the research landscape expands, adversarial attacks on models handling sequential data, such as Recurrent Neural Networks (RNNs) and Long Short-Term Memory networks (LSTMs), gain traction. Xu et al.~\cite{b32} delve into adversarial attacks on CNN+RNN-based sequential recognition (SR) tasks, focusing on scene text recognition (STR) and image captioning (IC). Their algorithm addresses limitations in existing attack methods for SR models, formulating optimization problems with novel objectives for efficient attacks on STR and IC models.

Fursov et al.~\cite{b33} contribute two innovative approaches for adversarial attacks on sequence data. The MCMC Attack employs a Monte Carlo search in the embedded sequence space, guided by an energy function based on the distance between initial and generated sequences. The CASCADA Attack utilizes differentiable sequential distance metrics and gradient optimization to generate adversarial examples, transforming the problem into a differentiable loss function.

In conclusion, adversarial attacks on neural networks pose a substantial challenge, with researchers continually developing new attack and defense methods. Machine learning practitioners must remain vigilant, staying abreast of these attacks and working to develop robust models capable of resisting them.

\section{A Running Example On RNN}
\begin{figure}[htbp]
\centerline{\includegraphics[width=0.5\textwidth]{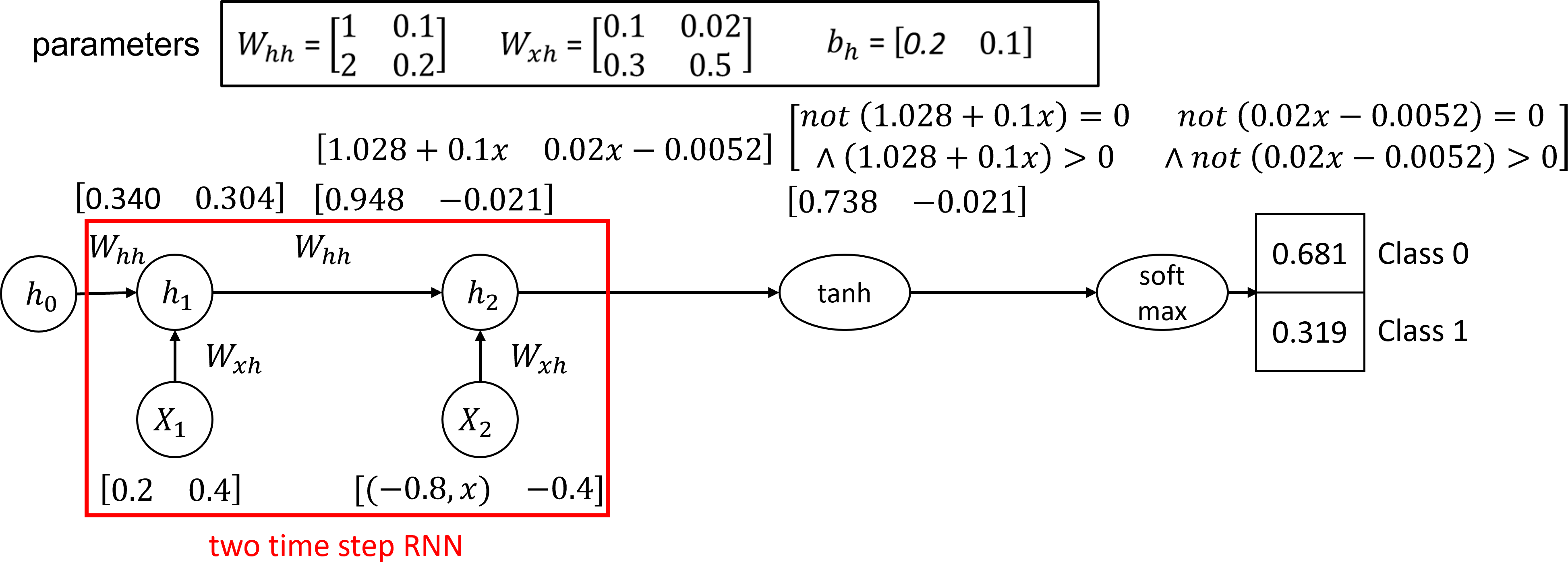}}
\caption{A simple RNN model.}
\label{rnn model}
\end{figure}

PyCT, a concolic testing tool, repeatedly executes the current test input to generate constraints corresponding to different branches in a tree $T$, representing the entire neural network. It solves the constraints in order in the Queue $Q$ to generate new test input that can alter the class against the initial input, resulting in an adversarial example. In the following, we use a simple network computation shown in Fig.~\ref{rnn model} as an example to illustrate how the algorithm works, including how $T$ and $Q$ are maintained. Fig.~\ref{rnn model} shows an input $X$ that takes a two-time-step SimpleRNN, tanh, and softmax to a two-class category. Assume that we select the second time step's first variable as a symbolic input, a variable that we can perturb its value.

\begin{figure}[htbp]
\centerline{\includegraphics[width=0.5\textwidth]{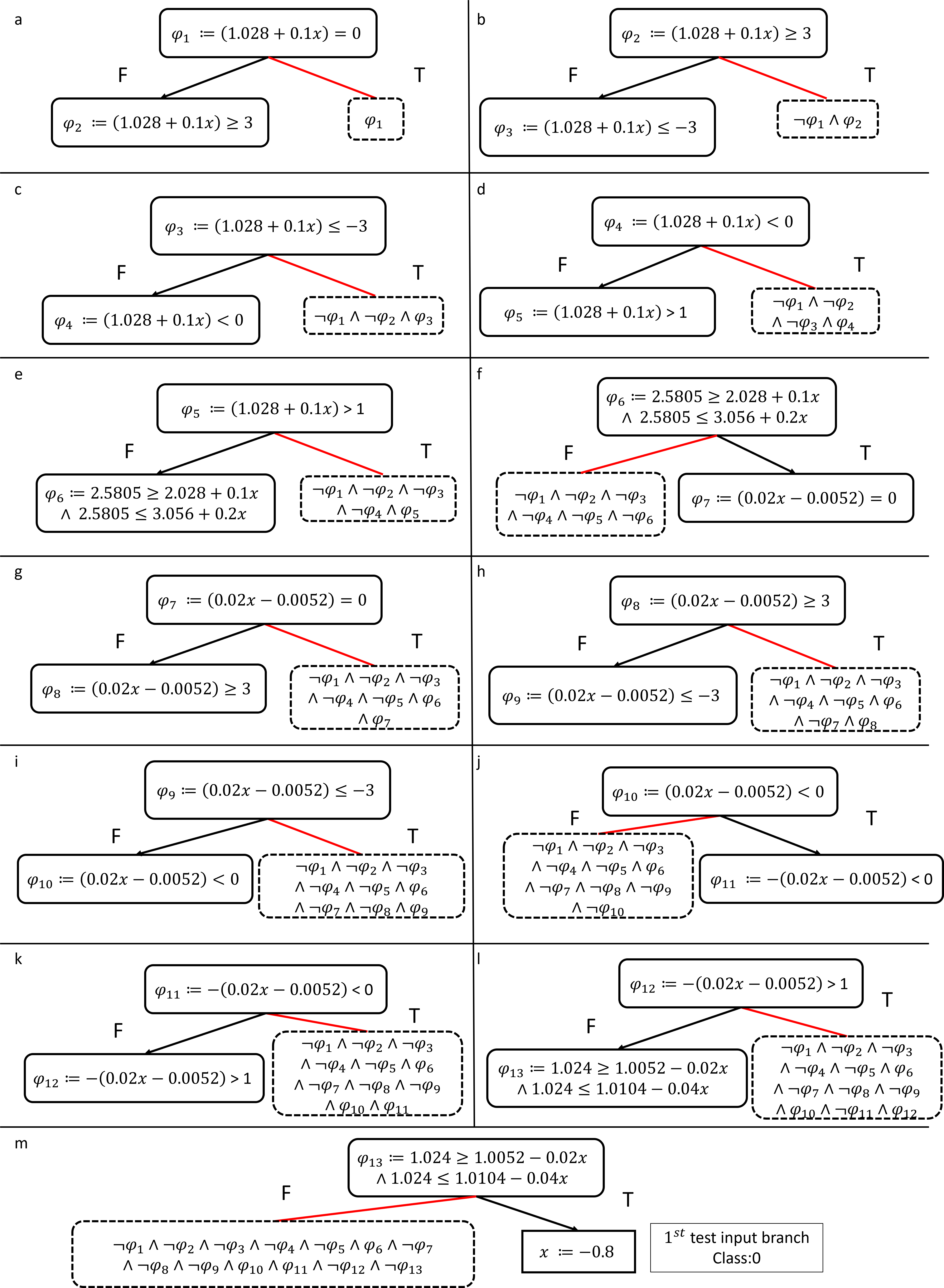}}
\caption{Tree $T$ after $1^{st}$ iteration.}
\label{tree_1}
\end{figure}

\begin{figure}[htbp]
\centerline{\includegraphics[width=0.5\textwidth]{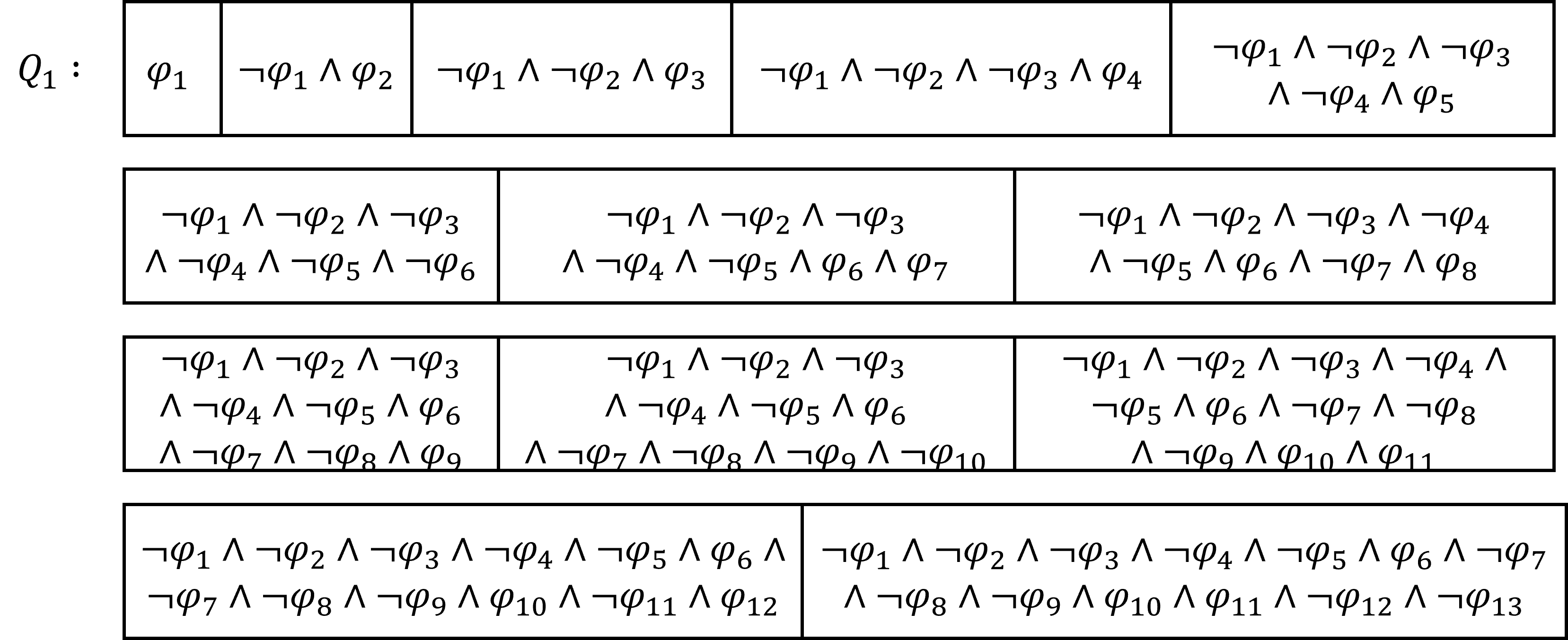}}
\caption{Queue $Q$ after $1^{st}$ iteration.}
\label{queue_1}
\end{figure}

In the first iteration, we provide a 2$\times$2 test input, with the selected variable initially set to -0.8. We create a concolic float $c_v$ = (-0.8, $x$) for the selected variable, and keep the other values as constant floats. Computing $h_1$ (the output of the first time step) gives [0.34, 0.304] after applying the formula $h_0\times W_{hh}+X_1\times W_{xh}+b_h$ = [(0$\times$1+0$\times$2)+(0.2$\times$0.1+0.4$\times$0.3)+0.2, (0$\times$0.1+0$\times$0.2)+(0.2$\times$0.02+0.4$\times$0.5)+0.1]. Then, we obtain $h_2$ using $h_1$, the input, and the parameters, resulting in a concrete value of [0.948, -0.021], and a symbolic expression of [1.028+0.1$\times x$, 0.02$\times x$-0.0052]. As PyCT passes through the tanh activation layer, it generates constraints. Since the concrete value of $h_2[0]$ is 0.948, which is not equal to 0, it will generate the constraint $1.028+0.1\times x = 0$ in the first 'if' condition. This constraint can be represented as $\phi_1$ and added to $Q_1$ for the next iteration's computation. Since $h_2[0]$ is not greater than or equal to 3, it proceeds to the second 'elif' condition, generating the constraint $\neg(1.028+0.1\times x = 0) \land (1.028+0.1\times x \geq 3)$, which can be represented as $\neg\phi_1 \land \phi_2$ and added to $Q_1$. Next, as $h_2[0]$ is not less than or equal to -3, it proceeds to the third condition, generating the constraint $\neg(1.028+0.1\times x = 0) \land \neg(1.028+0.1\times x \geq 3) \land (1.028+0.1\times x \leq -3)$, represented as $\neg\phi_1 \land \neg\phi_2 \land \phi_3$, and added to $Q_1$. In the final stages of the tanh operation, we substitute $1.028 + 0.1 \times x$ into our designed exponential function. Consequently, we enter the first 'if' statement of that function, which checks if $1.028 + 0.1 \times x < 0$. Since $0.948$ is positive, a new constraint is generated as $\neg\phi_1 \land \neg\phi_2 \land \neg\phi_3 \land \phi_4$ and added to $Q_1$. Moving on to the next 'elif' statement, which examines whether $1.028 + 0.1 \times x > 1$, as it doesn't satisfy this condition, it results in generating $\neg\phi_1 \land \neg\phi_2 \land \neg\phi_3 \land \neg\phi_4 \land \phi_5$ and adding it to $Q_1$. Finally, within the exp function, when the conditions $2.5805 \geq 2.028 + 0.1 \times x \land 2.5805 \leq 3.056 + 0.2 \times x$ which is simplified from the last statement of the exp function $(1 + x < math.exp(x) < 1 + 2x)$ satisfy, it generates the condition $\neg\phi_1 \land \neg\phi_2 \land \neg\phi_3 \land \neg\phi_4 \land \neg\phi_5 \land \neg\phi_6$ and adds it to $Q_1$. After completing the part for $h_2[0]$, PyCT will continue to generate new constraints for $h_2[1]$ at the tanh layer. Because $h_2[1]$ is -0.021, it will generate the constraint $\neg(1.028+0.1\times x = 0) \land \neg(1.028+0.1\times x \geq 3) \land \neg(1.028+0.1\times x \leq -3) \land \neg(1.028+0.1\times x < 0) \land \neg(1.028+0.1\times x > 1) \land ((2.5805 \geq 2.028 + 0.1 \times x) \land (2.5805 \leq 3.056 + 0.2 \times x)) \land (0.02\times x-0.0052 = 0)$, represented as $\neg\phi_1 \land \neg\phi_2 \land \neg\phi_3 \land \neg\phi_4 \land \neg\phi_5 \land \phi_6 \land \phi_7$, and added to $Q_1$. The process continues, generating constraints for the other conditions at the tanh layer and exp function then adding them to $Q_1$. After passing through the tanh layer, the output values are [0.738, -0.021]. Next, since the final softmax layer has no condition settings, the output values are directly obtained as [0.681, 0.319]. We can then determine that the final predicted class is 0, completing the generation of constraints for this iteration. The entire process follows the sequence from 'a' to 'm' in Fig.\ref{tree_1}, and the resulting $Q_1$ is shown in Fig.\ref{queue_1}. ($x$=-0.8, class: 0)

\begin{figure}[htbp]
\centerline{\includegraphics[width=0.5\textwidth]{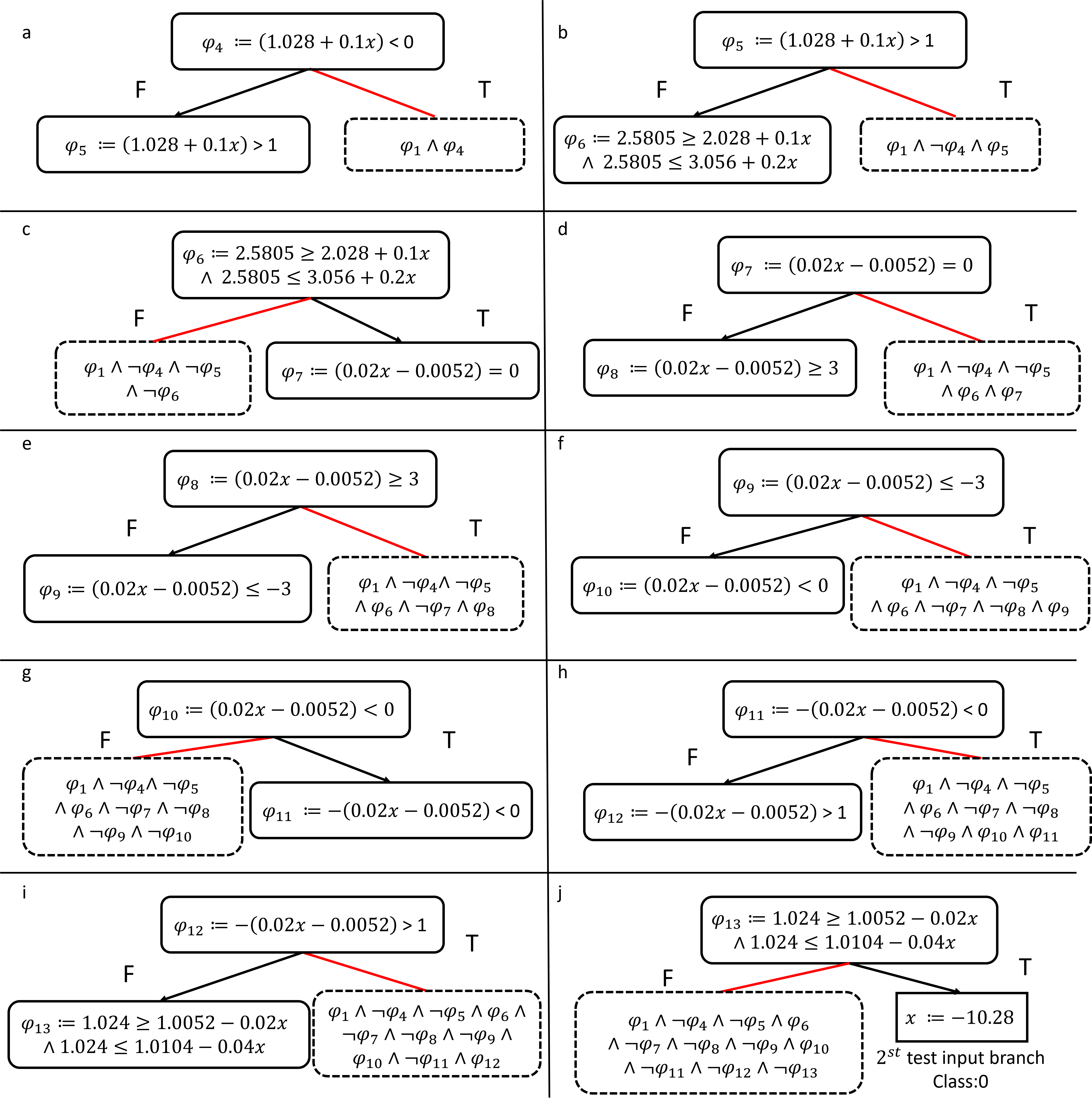}}
\caption{Tree $T$ after $2^{nd}$ iteration.}
\label{tree_2}
\end{figure}

\begin{figure}[htbp]
\centerline{\includegraphics[width=0.5\textwidth]{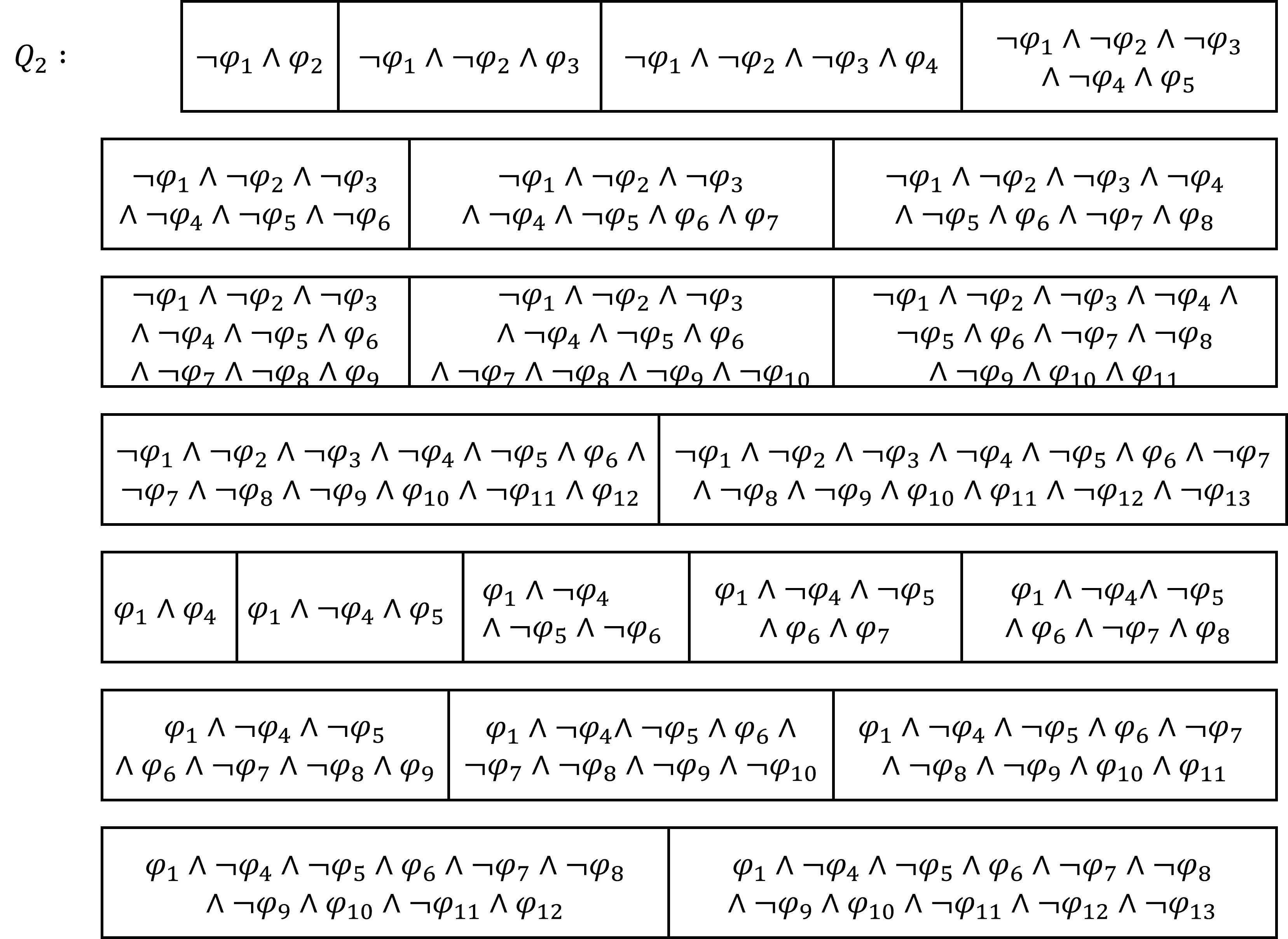}}
\caption{Queue $Q$ after $2^{nd}$ iteration.}
\label{queue_2}
\end{figure}

In the second iteration, we start by popping the formula $\phi_1$ from the front of $Q_1$. Since $\phi_1$ has a solution where $x=-10.28$, we can set the new initial value as the concolic float $c_v$=(-10.28, $x$). This leads to a concrete value of $h_2$ as [0, -0.211]. Similarly, we start with $h_2[0]$ passing through the tanh layer's conditions to generate constraints. This time, since $h_2[0]$ is equal to 0, we start generating constraint from exp function which starts from $1.028 + 0.1 \times x < 0$, $1.028 + 0.1 \times x > 1$ and then $2.5805 \geq 2.028 + 0.1 \times x \land 2.5805 \leq 3.056 + 0.2 \times x$. Therefore, PyCT continues to process $h_2[1]$ passing through the tanh layer's conditions. As $h_2[1]$ is -0.211, which is not equal to 0, it generates the constraint $\neg(1.028+0.1\times x < 0) \land \neg(1.028+0.1\times x > 1) \land ((2.5805 \geq 2.028 + 0.1 \times x) \land (2.5805 \leq 3.056 + 0.2 \times x)) \land (0.02\times x-0.0052 = 0)$, represented as $\phi_1 \land \neg\phi_4 \land \neg\phi_5 \land \phi_6 \land \phi_7$, and adds it to $Q_2$. Again, it proceeds through all other conditions in the tanh layer and generates constraints sequentially, adding them to $Q_2$. Finally, after passing through the tanh layer, the output values are [2.22, -0.2077], and through the last softmax layer, the output becomes [0.552, 0.448]. The predicted class is still 0, and this iteration ends. We have demonstrated PyCT's operation through two iterations in this example. The tree-like process followed in the second iteration is shown in Fig.\ref{tree_2}, from 'a' to 'j'. The resulting $Q_2$ is displayed in Fig.\ref{queue_2}. ($x$=-10.28, class: 0)

\section{Methodology}
\subsection{PyCT}
\begin{figure}[htbp]
\centerline{\includegraphics[width=0.5\textwidth]{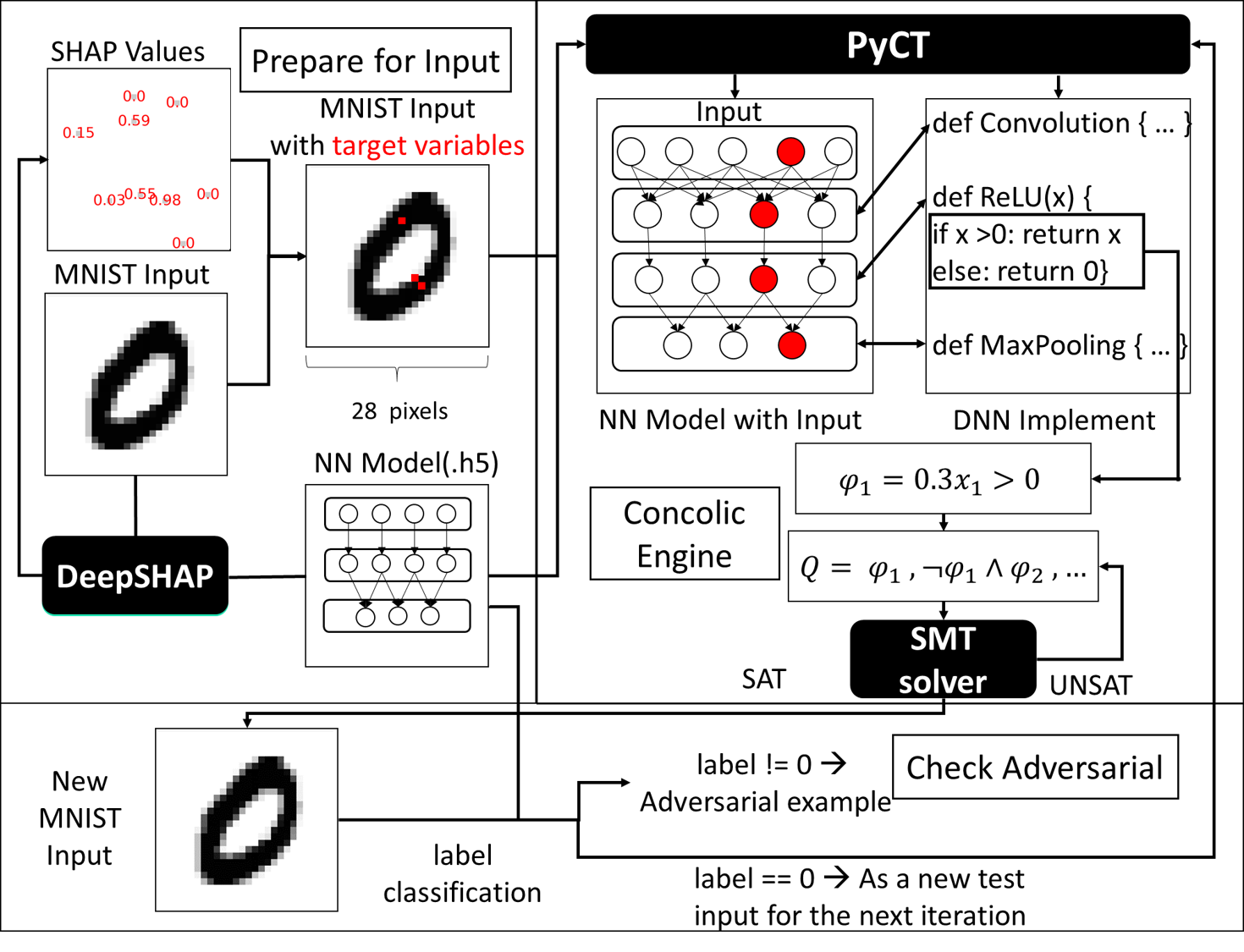}}
\caption{Process of utilizing PyCT.}
\label{PyCT structure}
\end{figure}

PyCT~\cite{b17} is a Concolic Testing tool that can generate adversarial examples using a program code and the initial test input for a neural network model. PyCT generates the adversarial example by selecting one or several pixels of the initial input as concolic variables. Then, when the concolic variables pass the conditional statements, PyCT generates several constraints that can reach different branches of the neural network model. Finally, PyCT iteratively selects a constraint from the queue to solve with an SMT solver in order to
explore new branches. If the constraint is UNSAT, the next constraint is poped out from the queue to be solved. The SMT solver returns the new input when the constraint is SAT. PyCT then checks whether it is an adversarial example (the inference result is different from the original one). If so, the process terminates. Otherwise, the input is used to conduct the concolic testing process and new constraints are added to the queue. The process continues until an adversarial example has been found or no constraints left to be solved.~\ref{PyCT structure} shows the process of utilizing PyCT.

We demonstrate the usage of PyCT for Deep Neural Network (DNN) models. Subsequently, PyCT generates constraints when the concolic variables pass through activation layers, and it solves the generated constraints, which can reach different branches in the neural network with SMT solver to generate test inputs, enabling the alteration of the predicted class output for adversarial example generation.

PyCT converts the target concolic variable into a concolic float type, where $c.val$ is a constant float, and $c.exp$ denotes its symbolic expression. We write $c = (f,x)$ for $c.val = f$ and $c.exp = x$. Specifically, each variable $v$ of the neural network corresponds to a concolic object $c_v$. When the concolic object $c_v$ passes through a neural network layer with conditional statements, it updates the value $c_v.val$ along with $v$, and $c_v.exp$ describes the relation between the current value of $v$ and the initial input values.

\subsection{The process of attacking DNN}
The process of generating adversarial examples for a DNN model using PyCT is depicted in~\ref{PyCT structure}, and it can be primarily divided into three main sections. The first section, named "Prepare for Input", is dedicated to the selection of target variables from the given DNN model and an initial test input. The second section, named "Concolic Engine", outlines how PyCT generates constraints to create potential adversarial test cases, thereby determining the success of the attack process. The final part, named "Check Adversarial", focuses on verifying whether the generated test cases qualify as adversarial examples. Each of these three sections will be elaborated on in the subsequent sections.

\subsubsection{Target variable selection policies}
The selection function for the target attribute plays a crucial role in our experiment. In PyCT, the newly generated test cases only change the values at the target variables to achieve a perturbation effect. We use two types of selection functions in our experiment: Random selection $S_{rnd}$ and SHAP value selection $S_{SHAP}$. Random selection entails randomly selecting a fixed number of attributes from all the attributes in the test input, each with the same probability.

On the other hand, the SHAP value selection involves choosing the attributes in the test input that have the most positive influence on the original predicted label, as indicated by the highest SHAP value determined by DeepSHAP~\cite{b18}.

DeepSHAP (SHapley Additive exPlanations)~\cite{b18} is a popular method for explaining a machine learning model by attributing the contribution of each feature to the output. SHAP values are based on game theory and provide a unique way to assign a value to each feature in a prediction by computing the average marginal contribution of that feature across all possible coalitions of features. We can use the SHAP values to identify the features that significantly impact model's output. By selecting these features, we can generate adversarial examples that specifically target these features and are more likely to cause the model to misclassify. Instead of generating adversarial examples for all possible combinations of features, we can use the SHAP values to identify the most critical features and only generate adversarial examples for these features. SHAP values can significantly reduce the number of queries required and improve the efficiency of the adversarial attack.

\subsubsection{Concolic Testing}
In Algorithm~\ref{alg:pyct_concolic}, we explore new branch conditions along the execution of the test input to expand Tree $T$ and path constraints queue $Q$. In Line 4 to 8, a new constraint that take the same prefix branches but different last one is generated and pushed to $Q$. Each formula in $Q$ specifies a new possible test input to take a different path (each takes the same prefix but differs in the last branch) of the current execution. The constraints are generated and inserted to explore all the branches identified in this iteration.

\begin{algorithm}[ht]
\caption{Concolic Testing on generating constraints}
\label{alg:pyct_concolic}
\begin{algorithmic}[1]
\Require Neural network $N$, initial input $X$, target concolic variable $c_v$
\Ensure Queue $Q$ with new test input constraints, and $\Phi$ the explored branch conditions on $c_v$ in $N(X)$ 
\Procedure{Execute}{$N$, $X$, $c_v$, $Q$, $T$}
\State $\varphi$ = $\varphi_0$ = constraint($T$) //Initialize the current constraints from T
\State $\Phi$ = \text{ConcolicExecution}($N$,$X$,$c_v$,$T$)
\ForAll{$\phi$ in $\Phi$}
\State $T$.append(constraintNode($\phi$))
\State $\varphi_{neg} \gets \varphi \land \neq\phi$
\State $\varphi \gets \varphi \land \phi$
\State $Q$.push($\varphi_{neg}$)
\EndFor
\State $T$.append(decisionNode($X$))
\State \textbf{return} $Q$, $T$
\EndProcedure
\end{algorithmic}
\end{algorithm} 

\subsubsection{Adversarial attack synthesis}
Algorithm~\ref{alg:pyct_oracle} summarizes the algorithm to check if the test cases generated from the given input ($X$) against the neural network ($N$) on a set of selected variables ($c_v$) is an adversarial example ($X'$) that changes the prediction class, i.e., $N(X')!=N(X)$ (Line 8), where only the values of the selected variables can be changed. This is done by leveraging the concolic execution (Line 11) iteratively on all models of constraints in $Q$ (Line 4 to 14). The constraint can be pop from the end as Stack (as Line 5) or the front as Queue (change $Q$.pop() to $Q$.pop(0) in Line 5). The Queue strategy adopts the breath first search, where PyCT explores inputs that turn at the first branches; the Stack strategy adopts the depth first search, where PyCT explores inputs that turn at the deepest branches first.

\begin{algorithm}[ht]
\caption{Adversarial attack synthesis}
\label{alg:pyct_oracle}
\begin{algorithmic}[1]
\Require Neural network $N$, initial input $X$, target concolic variable $c_v$
\Ensure New test input $X'$ generated from $X$ by PyCT
\Procedure{CheckAdversarial}{$N$, $X$, $c_v$}
\State Initialize $Q$, $T$ //Both are empty at the beginning
\State $Q$, $T \gets \text{Execute}(N,X,c_v,Q,T)$
\While{$Q$ is not empty}
\State $\varphi \gets Q$.pop()
\If{$\varphi$ is SAT}
\State $X' \gets \text{model}(\varphi)$
\If{$N(X') \neq N(X)$}
\State \textbf{return} $X'$, $T$
\Else
\State $Q, T \gets \text{Execute}(N,X',c_v,Q,T)$
\EndIf
\EndIf
\EndWhile
\State \textbf{return} $null$, $T$
\EndProcedure
\end{algorithmic}
\end{algorithm} 

\subsection{The implementation of DNN layer}
To generate constraints for the SMT solver, PyCT uses concolic variables to perform computations within the neural network. Performing these computations using concolic variables is not something that Keras is capable of doing. Below we discuss how we define and implement the Convolution layer, Max-Pooling layer, SimpleRNN layer, LSTM layer, and activation layer from Keras using only built-in Python functions within PyCT. The main concept behind implementing these neural network layer in PyCT is to use for-loops instead of using numpy for computing matrix multiplication as done in Keras.

\subsubsection{Activation Layer}
PyCT generates constraints when the concolic variables pass through activation layers such
as the Rectified Linear Unit (ReLU) layer, the Hyperbolic Tangent (tanh) layer and the Sigmoid layer, which DNN models commonly use. Algorithm~\ref{activation} shows how we implement these activation functions with the Python program. $T_\sigma$ ($T_{tanh}$) denotes the sigmoid (tanh) threshold for the extreme value. We add additional conditions to trigger different scenarios with the aim of imposing constraints on PyCT to explore additional branches related to specific scenarios. We aim to have PyCT explore different branches in order to investigate adversarial examples in greater details. For instance, the tanh layer transforms input values into output values within the range of -1 to 1. In the implementation, we have added branches to explore three extrem scenarios These conditions correspond to output values of 0 ($x=0$), 0.9951 ($x\geq3$, i.e., $T_{tanh}$=3), and -0.9951 ($x\leq-3$), respectively. This leads the SAT inputs of path constraints to explore different scenarios during the concolic testing process.

\begin{algorithm}
\caption{Activation layer Implementation}
\label{activation}
\begin{algorithmic}[1]
\Function{Sigmoid}{$x$}
  \If{$x == 0$}
    \State \Return $1.0 / (1.0 + \exp(-x))$
  \ElsIf{$x \geq T_\sigma$}
    \State \Return $1.0 / (1.0 + \exp(-x))$
  \ElsIf{$x \leq -T_\sigma$}
    \State \Return $1.0 / (1.0 + \exp(-x))$
  \Else
    \State \Return $1.0 / (1.0 + \exp(-x))$
  \EndIf
\EndFunction

\Function{Tanh}{$x$}
  \State $e_x \gets \exp(x)$
  \State $e_{-x} \gets \exp(-x)$
  \If{$x == 0$}
    \State \Return $(e_x - e_{-x}) / (e_x + e_{-x})$
  \ElsIf{$x \geq T_{tanh}$}
    \State \Return $(e_x - e_{-x}) / (e_x + e_{-x})$
  \ElsIf{$x \leq -T_{tanh}$}
    \State \Return $(e_x - e_{-x}) / (e_x + e_{-x})$
  \Else
    \State \Return $(e_x - e_{-x}) / (e_x + e_{-x})$
  \EndIf
\EndFunction

\Function{ReLU}{$x$}
  \If{$x < 0$}
    \State \Return $0$
  \Else
    \State \Return $x$
  \EndIf
\EndFunction
\end{algorithmic}
\end{algorithm}

\subsubsection{Convolutional Layer}
A convolutional layer is a fundamental component of Convolutional Neural Networks (CNNs) used for processing images and two-dimensional data. It is inspired by the biological visual system and is designed to capture local features in images, which are then used for advanced image classification and segmentation tasks. The basic building blocks of a convolutional layer are convolutional kernels, pooling layers, and fully connected layers. The convolution operation involves sliding a convolutional kernel (also known as a filter) over an input image and computing the weighted sum of the local region. The size and number of convolutional kernels are adjustable, allowing the model to automatically learn different levels of features. The output of a convolutional layer is referred to as a feature map, which represents mappings of various features in the input image. The mathematical expression governing the convolution operation within each output feature map is articulated as follows:
\begin{equation}
\begin{split}
    & O(i, j) = \sum_{row=0}^{m} \sum_{col=0}^{n} \sum_{dep=0}^{l} \\
    & I(i \cdot S + row, j \cdot S + col, dep) \cdot K(row, col, dep) + b
\end{split}
\end{equation}

Here, $O(i, j)$ denotes a position within the convolution result. Here, $I$ refers to the input image, $K$ represents the convolutional kernel, and $m$, $n$, and $l$ correspond to the height, width, and depth of the kernel. The stride size is denoted as $S$, and $(i, j)$ signifies a position within the output feature map. The indices of the convolutional kernel are represented as $(row, col, dep)$.

In Algorithm~\ref{cnn}, we present a detailed procedure for the implementation of a convolution operation. This convolution function takes parameters including the input tensor $I$, filter weights $W_f$, bias vector $b$, and stride $S$. The algorithm begins by obtaining the shapes of the input and filter, calculating the corresponding output shape, and initializing a zero vector $O$ for output (Lines 2 to 6).

Subsequently, three nested loops iterate over the depth ($k$), rows ($i$), and columns ($j$) of the output, as well as the depth, rows, and columns of the filter. Within the inner two loops, the algorithm performs element-wise multiplication of the input and filter over the corresponding regions, accumulating the results at the appropriate positions in the output tensor. Additionally, the bias term $b$ is added to the output of each filter (Lines 7 to 13). Once all loops are completed, the function returns the resulting output tensor $O$ (Line 14).

\begin{algorithm}
\caption{Convolutional Implementation}
\label{cnn}
\begin{algorithmic}[1]
\Function{Convolution}{Input as $I$, filter weights as $W_f$, bias as $b$, stride as $S$}
    \State $in\_s \gets I$.shape
    \State $f\_s \gets W_f$.shape
    \State $m, n, l \gets f\_s[1], f\_s[2], f\_s[3]$
    \State $out\_s \gets ((in\_s[0] - m) // S + 1, (in\_s[1] - n) // S + 1, f\_s[0])$ //$f_s[0]$ means the amount of filters
    
    \State Initialize $O$ to zero vectors of length $(out\_s)$
    \For{$k, i, j$ in product(range($out\_s[2]),$range($out\_s[0]),$
    \State \hspace{4em}range($out\_s[1])$}
        \For{$row, col, dep$ in product(range($i \cdot S, i \cdot S + m),$
        \State \hspace{4em}range($j \cdot S, j \cdot S + n),$range($0, l))$}
            \State $O[i][j][k] \gets I[row][col][dep] \cdot W_f[k][row - i \cdot S][col - j \cdot S][dep]$
        \EndFor
        \State $O[i][j][k] \gets O[i][j][k] + b[k]$
    \EndFor
    \State return $O$
\EndFunction

\end{algorithmic}
\end{algorithm}

Here, $F(i, j)$ is a position in the convolution result, $I$ is the input image, $K$ is the convolutional kernel, $S$ is stride size, $(i, j)$ is a position in the output feature map, $(m, n)$ are indices of the convolutional kernel.

\subsubsection{Pooling Layer}
The pooling layer typically follows a convolutional layer and is used to reduce the size of the feature map and computational cost. The most common pooling operations are max pooling and average pooling. Max pooling is particularly useful for achieving local invariance by selecting the maximum value within a region as the output. The mathematical expression for max pooling operation is as follows:
\begin{equation}
\begin{split}
    & O(i, j, k) = max\{I(row,col,k) \\
    & :i \cdot S \leq row < i \cdot S + m, j \cdot S \leq col < j \cdot S + n\}
\end{split}
\end{equation}

Here, $O(i, j)$ is a position in the pooling result, $I$ is the input feature map, $(i, j)$ is a position in the output feature map, $(m, n)$ are indices of the pooling region, and $s$ is the pooling stride. In MaxPooling, the pool window dimensions $(m, n)$ are typically set to match the pooling stride $s$. This enhances computational efficiency, ensures that each feature value is used only once, maintains consistent feature extraction, and facilitates control over the output feature map size. These benefits collectively contribute to efficient feature downsampling in convolutional neural networks. In Algorithm~\ref{maxpooling}, we also configure the values of $(m, n)$ to be equal to $s$.

CNNs typically consist of multiple stacked convolutional and pooling layers to gradually extract abstract features from images, which are then used for tasks such as image classification or object detection. This hierarchical feature extraction has made CNNs highly successful in image processing.

In Algorithm~\ref{maxpooling}, we illustrate the implementation of a MaxPooling layer in PyCT. The function takes an input tensor and a pooling stride size as parameters. It initializes various variables, such as the sizes of the input and output tensors, the values of $m$ and $n$, and creates an empty output tensor $O$ (Line 2 to 6).

Subsequently, we utilize nested loops to iterate through the rows, columns, and depth (channels) of the output tensor, employing the variables $row$, $col$, and $dep$ for iteration. In each iteration, we calculate the starting and ending positions of the pooling window on the input tensor (Line 10 to 13). These positions specify the location of the pooling window within the input tensor.

Following this, we use additional nested loops to collect all the values within each pooling window (Line 14 to 19). Finally, we determine the maximum value within each pooling window (Line 20 to 21). Once all iterations are completed, the function returns the output tensor $O$, which serves as input for the subsequent neural network layer (Line 25).

\begin{algorithm}
\caption{MaxPooling Implementation}
\label{maxpooling}
\begin{algorithmic}[1]
\Function{MaxPooling}{Input as $I$, Stride as $S$, Pool size as $p$}
  \State $in\_s \gets I$.shape
  \State $m \gets p[0]$
  \State $n \gets p[1]$
  \State $out\_s \gets ((in\_s[0] - m )// S + 1, (in\_s[1] - n )// S + 1, in\_s[2])$
  \State Initialize $O$ to zero vectors of length $(out\_s)$
  \For{$i, j, k$ in product(range($out\_s[0]),$range($out\_s[1]),$
  \State \hspace{8em}range($out\_s[2])$)}
    \State $window \gets []$
    \For{$row$ in range$(i \cdot S, i \cdot S + m)$}
      \For{$col$ in range$(j \cdot S, j \cdot S + n)$}
        \State $window$.append$(I[row][col][k])$
      \EndFor
    \EndFor
    \State $max\_val \gets \max(window)$
    \State $O[i][j][k] \gets max\_val$
  \EndFor
  \State \Return $O$
\EndFunction
\end{algorithmic}
\end{algorithm}

\subsubsection{Recurrent Neural Network (RNN)}
A recurrent neural network (RNN) is an artificial neural network that processes sequential data, such as time series or natural language text. Unlike feedforward neural networks, which process input data in a fixed order and do not have any memory of previous inputs, RNNs have a 'memory' that allows them to consider previous inputs when processing current ones.

The main idea behind RNNs is to use feedback loops that connect the outputs from previous time steps to the inputs of the current time step. This allows the network to maintain a hidden state that captures information about the previous inputs. The model updates the hidden state at each time step based on the current input and the previous hidden state. The following equation describes the actual computation of how RNN takes previous memory into account:
\begin{equation}
    h_{t} = tanh(x_{t} \cdot W_{xh} + h_{t-1} \cdot W_{hh} + b_h)
\end{equation}

The implementation of the RNN layer is presented in Algorithm~\ref{rnn}. The process begins with the creation of an empty tensor to hold the output hidden state $h_t$ (Line 2). Subsequently, a nested loop is initiated to calculate each value of the output hidden state sequentially.

Before computing the hidden state value, a temporary variable $h$ is initialized for each iteration (Line 4). Then, two distinct for-loops are employed: one for multiplying the input $x$ with the weights between input and hidden state $W_{xh}$ (Line 5 to 7), and another for-loop for multiplying the last hidden state $h_{t-1}$ with the weights between the last and current hidden state $W_{hh}$ (Line 8 to 10).

Upon adding the bias $b_h$ and applying the hyperbolic tangent (tanh) function, the hidden state value for the current iteration is obtained (Line 11 to 12). After completing all iterations, the function returns the final hidden state $h_t$ (Line 14).

\begin{algorithm}
\caption{RNN Implementation}
\label{rnn}
\begin{algorithmic}[1]
\Function{RNN}{input as $x$, last hidden state as $h_{t-1}$, weight as $W_{xh}$ and $W_{hh}$, bias as $b_h$, output shape as $units$}
    \State Initialize $h_t$ to zero vectors of length $units$
    \For{$i$ in range($units$)}
      \State $h \gets 0$
      \For{$j$ in range($units$)}
        \State $h \mathrel{+}= h_{t-1}[j] \cdot W_{hh}[j][i]$
      \EndFor
      \For{$j$ in range(len($x$))}
        \State $h \mathrel{+}= x[j] \cdot W_{xh}[j][i]$
      \EndFor
      \State $h \mathrel{+}= b_h[i]$
      \State $h_t[i] \gets$ tanh($h$)
    \EndFor
    \State \Return $h_t$
\EndFunction
\end{algorithmic}
\end{algorithm}

\subsubsection{Long Short-Term Memory (LSTM)}
LSTM is an acronym for Long Short-Term Memory, a type of recurrent neural network (RNN) architecture specifically designed to address the issue of vanishing gradients in traditional RNNs.

LSTM networks consist of LSTM cells linked together to form a sequence. Each LSTM cell has three gates (input, output, and forget) and a cell state that allows it to selectively remember or forget past inputs based on their relevance to the current task.

The forget gate $f$, controlled by a sigmoid function ($\sigma$), determines whether previously obtained information $C_{t-1}$ should pass through, based on an $f_t$ value between 0 and 1. We can express the computation of $f$ as follows:
\begin{equation}
f_t = \sigma(W_f \cdot [h_{t-1}, x_t] + b_f)
\end{equation}

An LSTM model's input gate $i$ determines whether to store information in the cell state. It consists of two steps: firstly, a sigmoid layer ($\sigma$) decides which data to update, and secondly, a tanh layer generates new information $\widetilde C_t$ that can be added to the cell state. We can express the computation of $i$ and $\widetilde C_t$ as follows:
\begin{equation}
i_t = \sigma(W_i \cdot [h_{t-1}, x_t] + b_i)
\end{equation}
\begin{equation}
\widetilde{C_t} = tanh(W_C \cdot [h_{t-1}, x_t] + b_C)
\end{equation}

After obtaining values from the input and forget gates, we can calculate the current cell state $C_t$ for the next time step (cell). We can express the computation of $C_t$ as follows:

\begin{equation}
C_t = f_t \times C_{t-1} + i_t \times \widetilde{C_t}
\end{equation}

The output gate $o$ controls the amount of information to be output. Firstly, an initial output is produced by a sigmoid layer, which scales $C_t$ to the range of [-1, 1] using tanh. Finally, the final output $h_t$ is obtained by element-wise multiplication of the initial output and the output obtained from the sigmoid layer. We can express the computations of $o$ and $h_t$ as follows:
\begin{equation}
o_t = \sigma(W_o \cdot [h_{t-1}, x_t] + b_o)
\end{equation}
\begin{equation}
h_t = o_t \times tanh(C_t)
\end{equation}

The ability of LSTM networks to selectively retain or discard past inputs makes them particularly valuable for tasks that require long-term dependencies, such as speech recognition, language translation, and image captioning.

We illustrate the LSTM layer in Algorithm~\ref{lstm}. Initially, four empty tensors $i, f, o, \widetilde C_t$ are created to store the values of the three gates and the candidate value, where $i$ represents the input gate (Line 2).

Subsequently, a nested loop is initiated to calculate the values of the three gates and the candidate value sequentially. Within the nested for-loop, two distinct for-loops are employed: one for multiplying the input $x$ with the weights between the input and hidden state $W$ (Line 4 to 9), and another for-loop for multiplying the last hidden state $h_{t-1}$ with the weights between the last and current hidden state $U$ (Line 10 to 15). The nested for-loop concludes with the addition of biases $b$ for the four tensors: $i, f, o, \widetilde C_t$ (Line 10 to 15).

After computing the values of the three gates and the candidate value, two empty tensors are initialized for the cell state and hidden state, which will be returned by this function (Line 21). The process culminates in the calculation of the cell state $C_t$ and the hidden state $h_t$ (Line 22 to 25).

\begin{algorithm}
\caption{LSTM Implementation}
\label{lstm}
\begin{algorithmic}[1]
\Function{LSTM}{input as $x$, last hidden state as $h_{t-1}$, last cell as $C_{t-1}$, weight as $W$ and $U$, bias as $b$, and output shape as $units$}
    \State Initialize $i, f, o, \widetilde C_t$ to zero vectors of length $units$
    \For{$j$ in range($units$)}
      \For{$k$ in range(len($x$))}
        \State $i[j] \mathrel{+}= x[k] \cdot W_i[k][j]$
        \State $f[j] \mathrel{+}= x[k] \cdot W_f[k][j]$
        \State $o[j] \mathrel{+}= x[k] \cdot W_o[k][j]$
        \State $\widetilde C_t[j] \mathrel{+}= x[k] \cdot W_C[k][j]$
      \EndFor
      \For{$l$ in range($units$)}
        \State $i[j] \mathrel{+}= h_{t-1}[l] \cdot U_i[l][j]$
        \State $f[j] \mathrel{+}= h_{t-1}[l] \cdot U_f[l][j]$
        \State $o[j] \mathrel{+}= h_{t-1}[l] \cdot U_o[l][j]$
        \State $\widetilde C_t[j] \mathrel{+}= h_{t-1}[l] \cdot U_C[l][j]$
      \EndFor
      \State $i[j] \mathrel{+}= b_i[j]$
      \State $f[j] \mathrel{+}= b_f[j]$
      \State $o[j] \mathrel{+}= b_o[j]$
      \State $\widetilde C_t[j] \mathrel{+}= b_C[j]$
    \EndFor
    \State Initialize $C_t$ and $h_t$ as zero vectors of length $units$
    \For{$j$ in range($units$)}
      \State $C_t[j] \mathrel{=}$ $\sigma$($f[j]$) $\cdot C_{t-1}[j] +$ $\sigma$($i[j]$) $\cdot$ tanh($\widetilde C_t[j]$)
      \State $h_t[j] \mathrel{=}$ $\sigma$($o[j]$) $\cdot$ tanh($C_t[j]$) //$\sigma$ represent sigmoid layer
    \EndFor
    \State \Return $h_t$ and $C_t$
\EndFunction
\end{algorithmic}
\end{algorithm}

\subsubsection{The Nature Exponential Function}
math.exp($x$) is the exponential function $e^x$ that is widely used for activation functions such as sigmoid and hyperbolic tangent (tanh). Facing such non linear function results in the downgrade of our concolic testing, where the concolic variables are downgraded to the concrete constant and no more symbolic expressions or constraints associated with them are generated. We add additional assertion (Line 8 in Algorithm~\ref{exp}) on $x$ before we return the value of
math.exp($x$). In this case, we can still generate a final condition on $x$ before it becomes a constant. The assertion states that $1+x < \text{math.exp}(x) < 1+2x$. This restricts the current value on $x$ and generates constraints to explore
new input that is not within this range.

Next, to establish the range $1+x < \text{math.exp}(x) < 1+2x$. In Algorithm~\ref{exp}, prior to the assertion $1+x < \text{math.exp}(x) < 1+2x$, we first define the range of $x$ as $0<x<1$ (Lines 2 to 5). Within this range, we can initially prove that $1+x < \text{math.exp}(x)$. This is because the exponential function $e^x$ is derived from the Taylor expansion, which can be expressed as $1 + x + \frac{x^2}{2!} + \frac{x^3}{3!} + \frac{x^4}{4!} + \ldots = \sum_{n=0}^{\infty} \frac{x^n}{n!}$, where 1 and $x$ represent the first two terms in the Taylor expansion. Since the subsequent terms are all positive due to the range $0<x<1$ and the addition of positive values, the value of $e^x$ exceeds $1+x$ for $0<x<1$.

When considering $\text{math.exp}(x) < 1+2x$, we first aim to prove that for a positive integer $n$, the $n$th term $\frac{x^n}{n!}$ is greater than the sum of the subsequent terms from $n+1$ to infinity, which can be represented as $\sum_{k=n+1}^{\infty} \frac{x^k}{k!}$. To achieve this, we initially express $\sum_{k=n+1}^{\infty} \frac{x^k}{k!}$ in the form of $x^n/n!$, as follows: $\frac{x^n}{n!}\sum_{k=n+1}^{\infty} \left(\frac{x^{k-n}}{k!/n!}\right)$.

\begin{lemma}
For all $n \geq 1$ and $0 < x < 1$:
\[
\frac{x^n}{n!} \geq \sum_{k=n+1}^{\infty} \frac{x^k}{k!}
\]
\end{lemma}

\begin{corollary}
For all $n \geq 1$ and $0 < x < 1$:
\[
1 + \ldots + \frac{x^n}{n!} < \exp(x) < 1 + \ldots + \frac{x^n}{n!} + \frac{x^n}{n!}
\]
\end{corollary}

Subsequently, since $0<x<1$, and $n$ is always a positive integer, it is evident that $x/n$ is greater than $x/(n+1)$, $x/(n+2)$, and so on. This allows us to establish $\frac{x^n}{n!}\sum_{k=n+1}^{\infty} \left(\frac{x^{k-n}}{k!/n!}\right) < \frac{x^n}{n!}\sum_{k=n+1}^{\infty} \left(\frac{x}{n}\right)^{k-n}$.

Now, in order to complete our proof, we need to demonstrate that $\sum_{k=n+1}^{\infty} \left(\frac{x}{n}\right)^{k-n} < 1$. This is achieved through an understanding of geometric series, and as 'n' approaches infinity, we can derive $\sum_{k=n+1}^{\infty} \left(\frac{x}{n}\right)^{k-n} = \frac{x/n}{1-{x/n}}$. Consequently, we can conclude that for values of $x$ such that $n \geq 2 \land 0<x<1$, the inequality $\frac{x^n}{n!} > \sum_{k=n+1}^{\infty} \frac{x^k}{k!}$ will always hold.

To establish the lemma for the base case, consider when $n = 1$. Setting $n = 2$, we observe that $\frac{x^2}{2!} > \sum_{k=2+1}^{\infty} \frac{x^k}{k!}$. Given that $0 < x < 1$, we can infer that $x > x^2$. Consequently, we demonstrate that even for $n = 1$, $x > \sum_{k=2}^{\infty} \frac{x^k}{k!}$. This successively proves that, for $n \geq 1$ and $0 < x < 1$, the lemma $\frac{x^n}{n!} \geq \sum_{k=n+1}^{\infty} \frac{x^k}{k!}$ consistently holds.

In our exponential function, we consider the case when $n = 1$ and assume $0 < x < 1$. This allows us to conclude that $\text{math.exp}(x) < 1+x+x$ based on the inequality $x > \sum_{k=2}^{\infty} \frac{x^k}{k!}$. Consequently, we confidently assert that $1+x < \text{math.exp}(x) < 1+2x$.

\begin{algorithm}
\caption{exp Implementation}
\label{exp}
\begin{algorithmic}[1]
\Function{\textbf{exp}}{$x$}
  \If{$x < 0$}
    \State \Return $1/exp(-x)$
  \ElsIf{$x > 1$}
    \State \Return $exp(x/2) \cdot exp(x/2)$
  \Else
    \State \textbf{assert} $1 + x < \text{math.exp}(x) < 1 + 2x$
    \State \Return $\text{math.exp}(x)$
  \EndIf
\EndFunction
\end{algorithmic}
\end{algorithm}

\section{Experiment}
\begin{enumerate}[label=\textbf{RQ\arabic*}:]
\item Can concolic testing imply on NN branches?
\item Does the order of branches affect prediction results?
\item Does picking the target position with DeepSHAP affect the attacking performance on PyCT?
\item How is the performance on solving different sizes of constraints with PyCT?
\item Is PyCT competitive to the state of the art NN constraint solver?
\end{enumerate}

\subsection{Dataset}
\subsubsection{Mnist}
The MNIST dataset is a well-known collection of handwritten digits, widely employed in machine learning and computer vision research. It comprises 28x28 pixel grayscale images representing digits from 0 to 9. In our experiment, we not only utilize the MNIST dataset with CNN models but also with RNN/LSTM models for the purpose of digit classification from 0 to 9.

\subsubsection{Trading Strategy Dataset}
This dataset is a custom creation of ours. We collected daily stock trading data from Microsoft for 20 years, spanning from April 15, 2003, to April 14, 2023. This dataset includes opening prices, closing prices, highest prices, and lowest prices. To create our dataset, we segmented this data into 20-day chunks.

Next, we devised a simple stock trading strategy: sell when there is a consecutive three-day increase in stock prices, and buy when there is a consecutive three-day decrease in stock prices. All other days remained inactive. Each data chunk was labeled with one of these three categories, and for each day within the chunk, we added 20 columns representing which day it was.

Following this, we randomly selected 80\% of the data for our training dataset, allocated 10\% for validation, and reserved the remaining 10\% as our test dataset (which also serves as the target for attacks).

\subsubsection{Internet Movie Database(IMDb)}
The Internet Movie Database (IMDb) is a widely recognized online database that offers information about movies, TV shows, actors, and other entertainment-related content. Users can access an extensive repository of reviews, ratings, and details for a broad spectrum of films and television programs. In our experiment, we will initially pass the embedding layer of the LSTM model based on TestRNN~\cite{b13}. The resulting 500x32 vectors will serve as the input dataset for IMDb in PyCT.

\subsection{Model}
\subsubsection{Simple CNN Model(CNN$\_$678)}
We first introduce a simple CNN model that is trained using the MNIST database including a set of 60,000 grayscale images of size 28x28 pixels. The goal of this model is to classify the images into different categories. It follows a sequential architecture and includes various layers. The model starts with a convolutional layer that applies 2 filters with a kernel size of 3x3 to the input images. ReLU activation functions are applied after the convolutional layer to introduce non-linearity to the model. Next, a max pooling layer with a pool size of 2x2 is applied to reduce the spatial dimensions of the feature maps. The process is repeated with another convolutional layer that also applies 2 filters with a kernel size of 3x3. ReLU activation functions follow the convolutional layer. Another max pooling layer is then applied to further reduce the spatial dimensions. The output from the pooling layers is flattened into a vector after passing through a flatten layer, which is then passed through a dense layer with 10 units. This dense layer applies a linear transformation to the input and introduces non-linearity using the ReLU activation function. Finally, the last dense layer are used to process the extracted feature information to get the classification
result. This model is a simple CNN model with only 678 parameters. The model achieves 96.28\% accuracy in the default MNIST test dataset (10 000 samples).

\subsubsection{Complex CNN Model(CNN$\_$2418)}
The CNN model I'm going to introduce is trained using the MNIST database, which consists of a set of 60,000 grayscale images of size 28x28 pixels. The goal of this model is to classify the images into different categories. It follows a sequential architecture and includes various layers. The model starts with a convolutional layer that applies 4 filters with a kernel size of 3x3 to the input images. ReLU activation functions are applied after the convolutional layer to introduce non-linearity to the model. Next, a max pooling layer with a pool size of 2x2 is applied to reduce the spatial dimensions of the feature maps. The process is repeated with another convolutional layer that also applies 4 filters with a kernel size of 3x3. ReLU activation functions follow the convolutional layer. Another max pooling layer is then applied to further reduce the spatial dimensions. The output from the pooling layers is flattened into a vector after passing through a flatten layer, which is then passed through a dense layer with 20 units. This dense layer applies a linear transformation to the input and introduces non-linearity using the ReLU activation function. Finally, the last dense layer are used to process the extracted feature information to get the classification
result. This model is a more complex CNN model with 2418 parameters. The model achieves 98.76\% accuracy in the default MNIST test dataset (10 000 samples).

\subsubsection{SimpleRNN Model(RNN)}
We use the MNIST database of 60,000 grayscale images of size 28x28 pixels to train a SimpleRNN model with two layers. The first layer in the model is a SimpleRNN layer with 32 units. This layer takes input sequences with a shape of (28, 28), where the first 28 represents the time steps, and the second 28 represents the number of features in each step. In this case, the input sequences correspond to rows of the input images. The first SimpleRNN layer encodes each input image as a row vector of shape (28, 128). The SimpleRNN layer processes the input sequences and captures temporal dependencies within the rows of the images. Then, we pass the output from the first SimpleRNN layer to the second layer, a Dense layer with ten units. This layer applies a linear transformation to the input and produces the classification result. This model is a simple RNN model with only 2282 parameters. The model achieves 96.5\% accuracy in the default MNIST test dataset (10 000 samples).

\subsubsection{LSTM Model(LSTM)}
We will introduce an LSTM model we trained using the MNIST database, which contains a set of 60,000 grayscale images of size 28x28 pixels. We utilized this dataset to train an LSTM model with two layers. The model utilizes an LSTM layer with 16 units, followed by a Dense layer with 10 units for classification. This model is a more complex model with 3050 parameters. The model achieves 97.85\% accuracy in the default MNIST test dataset (10 000 samples).

\subsubsection{Trading Strategy LSTM model(Stock)}
We utilized the Trading Strategy training dataset to train a simple LSTM model consisting of four layers. The initial layer in the model is a Dense layer with 8 units. This layer takes input sequences with 20 time steps (equivalent to 20 days) along with the daily stock price information at each step. Subsequently, the output from the first Dense layer is forwarded to the second layer, which is an LSTM layer with 8 units designed to capture the final hidden state information across the ten time steps. This is followed by another Dense layer with 16 units. Finally, the model passes through a Dense layer with 3 units, applying a linear transformation to the input and producing the classification result. This LSTM model is relatively simple, comprising only 939 parameters. The model achieved an accuracy of 88.04\% on the remaining test dataset, which consisted of 502 samples.

\subsubsection{IMDb LSTM model(IMDb)}
The IMDb LSTM model is derived from TestRnn~\cite{b16}, and we extract the embedding layer through which our dataset has passed. The input to this model is a vector of size 16,000. The model utilizes an LSTM layer with 100 units, followed by a Dense layer with 1 unit for classification. With a total of 53,301 parameters, the model achieves an accuracy of 85.09\% on the IMDb test dataset, which comprises 25,000 samples.

\subsection{Experimental Setup}
Our experiments targeted Convolutional Neural Network (CNN) models for classifying the MNIST dataset. For each image, we set the attack timeout to half an hour and performed attacks using the first 100 images from the test dataset. Additionally, for Recurrent Neural Network (RNN) and Long Short-Term Memory (LSTM) models for MNIST classification, as well as an LSTM model for stock trading strategies, the attack timeout was set to one hour. Attacks were conducted using either the first 100 images from the test dataset or 502 consecutive daily stock data points, spanning 20 days. In each attack, we employed a queue or stack-based constraint solving order and one of the two target pixel picking strategies: random or SHAP. In total, four combinations of attack strategies were employed.

In all our attack experiments on MNIST, we will initiate attacks by targeting a single pixel for all test cases. Subsequently, the failed test cases from the initial attack will undergo attacks on four pixels. This sequential process will continue in the order of 1, 4, 8, 16, 32, and so on until we reach 32, at which point the attacks will conclude. However, for stock trading strategy and IMDb, the attack sequence will follow the order of 1, 2, 3, 4, 8.

\subsection{Constraints on NN model(RQ1)}
In Table~\ref{tab1}, we compiled and analyzed the records of successful attacks for different models at varying numbers of target pixels. The success rate of attacks (ATK) is presented, accumulated in the order of pixel quantities: 1, 4, 8, 16, 32 or 1, 2, 3, 4, 8. Additionally, we computed the average quantities of satisfactorily generated new test cases (SAT), unsatisfactorily generated new test cases (UNSAT), and the average number of constraints generated per iteration for each test case (gen constraints). Finally, the average time required for each test case to achieve a successful attack (time) is also presented. Since all attacks on the Stock Trading model with pixel quantity equal to 8 resulted in failure, we have replaced the statistical row for this configuration with records from unsuccessful attack test cases.

The findings indicate a positive correlation between the quantity of target pixels and the generated constraints across five models, excluding the IMDb model. As the number of target pixels increases, there is a corresponding escalation in the count of generated constraints. We hypothesize that the limited constraint data generated for IMDb is attributed to the excessively short timeout settings. Consequently, the quantity of generated constraints does not increase proportionally with the higher selection of target pixels. To validate our conjecture, we conducted an extended duration attack experiment on the IMDb model. 

The subsequent experiment utilizes a queue-based constraint solving order and SHAP values for picking target pixels to launch attacks on the IMDb model. We tested with 1, 10, 100, and 1000 target pixels, each with a timeout of 12 hours, utilizing the same seven movie review test inputs. The results, as outlined in Table~\ref{tab0}. we calculated the average counts of successfully generated new test cases (SAT), unsuccessfully generated new test cases (UNSAT), and the average number of constraints generated per iteration for each test case (gen cstr). Ultimately, we determined the average time required for each individual forward pass (F time), and the statistic indicate that when choosing 100 or 1000 target pixels, the model did not successfully complete a single forward pass within the 12-hour timeframe for any of the seven test cases. With 10 target pixels, the model successfully completed the forwarding process only 18 times which is calculated by multiplying the average quantity of successfully generated new test cases (SAT) (2) by the number of test cases with successful forward passes (6), and then adding the initial count of test cases with successful forward passes (6) out of the seven test inputs, averaging 20798 seconds for each forward pass. Notably, one test case failed to forward even once within the 12-hour timeframe. In the end, with 1 target pixel, the model successfully completed the forwarding process 117 times. This count is calculated in the same manner as with 10 target pixels for the seven test inputs, resulting in an average forwarding time of 890 seconds. It is noteworthy that in these 28 test cases, there were no instances of successful attacks. Due to the difficulty of achieving successful attacks even with an extended timeframe, we opted to focus on attacking a subset of test inputs within a one-hour timeframe. This approach allows us to statistically analyze the forwarding and solving performance across various attacking processes.

\textbf{Answer to RQ1:} Based on the above statistics, it can be observed that both the model complexity and the quantity of target pixels affect the number of generated constraints. This indicates that in concolic testing tools like PyCT, constraints can indeed be generated for different branches of the NN model.

\begin{table*}
\centering
\caption{Amount of constraints with different model(Queue)}
\begin{tabularx}{0.85\textwidth}{lccccccccccccc}
\toprule
\textbf{} & \multicolumn{5}{c}{\textbf{CNN$\_$678}} & \multicolumn{4}{c}{\textbf{CNN$\_$2418}} \\
\cmidrule(r){2-6} \cmidrule(r){7-11}
\textbf{Pixel} & \textbf{ATK(\%)} &\textbf{SAT} & \textbf{UNSAT} & \textbf{gen constraints} & \textbf{time (s)} & \textbf{ATK(\%)} &\textbf{sat} & \textbf{unsat} & \textbf{gen constraints} & \textbf{time (s)} \\
\midrule
1 & 4\% & 1 & 6.5 & 115.38 & 6.85 & 1\% & 1 & 3 & 214 & 98.43 \\
4 & 10\% & 10.33 & 101 & 189.41 & 302.91 & 8\% & 1.14 & 2.43 & 536.53 & 289.96 \\
8 & 25\% & 25.27 & 339 & 255.15 & 417.67 & 18\% & 1.1 & 1.7 & 730.29 & 348.37 \\
16 & 33\% & 18.88 & 142.75 & 440.75 & 602.25 & 46\% & 1.57 & 9.43 & 1097.99 & 462.64 \\
32 & 53\% & 4.7 & 10.8 & 713.84 & 377.27 & 73\% & 1.07 & 5.44 & 1596.68 & 722.31 \\
\midrule
\textbf{} & \multicolumn{5}{c}{\textbf{RNN}} & \multicolumn{4}{c}{\textbf{LSTM}} \\
\cmidrule(r){2-6} \cmidrule(r){7-11}
1 & 1\% & 1 & 91 & 284 & 82.72 & 1\% & 1 & 0 & 402 & 5.64 \\
4 & 9\% & 2 & 54.38 & 605.88 & 116.3 & 11\% & 4.1 & 145.8 & 521.8 & 178.33 \\
8 & 27\% & 9.83 & 440.78 & 885.15 & 886.05 & 28\% & 2.18 & 71.59 & 1275.77 & 381.25 \\
16 & 45\% & 3.56 & 142.56 & 1893.21 & 757.16 & 44\% & 2.75 & 52.31 & 2166.55 & 932.7 \\
32 & 62\% & 1.06 & 25.94 & 3297.09 & 762.43 & 60\% & 1.06 & 8.44 & 5080.55 & 1162.36 \\
\midrule
\textbf{} & \multicolumn{5}{c}{\textbf{Stock}} & \multicolumn{4}{c}{\textbf{IMDb}} \\
\cmidrule(r){2-6} \cmidrule(r){7-11}
1 & 83.47\% & 5.67 & 123.78 & 65.36 & 82.82 & 0\% & 1.96 & 374.57 & 1456.19 & 3600\\
2 & 92.63\% & 39.3 & 1195.12 & 113.57 & 1057.01 & 0\% & 1.1 & 180.9 & 1634.1 & 3600\\
3 & 95.62\% & 40.87 & 1192.67 & 215.94 & 1834.28 & 0\% & 1.12 & 138.88 & 1773.25 & 3600\\
4 & 97.01\% & 34.43 & 1158.43 & 291.73 & 1898.25 & 0\% & 0.93 & 93.52 & 1660.84 & 3600\\
8 & 97.01\% & 42 & 1040.33 & 669.86 & 3600 & 0\% & 1.27 & 75 & 1611.32 & 3600\\
\bottomrule
\end{tabularx}
\label{tab1}
\end{table*}

\begin{figure}[htbp]
\centerline{\includegraphics[width=0.5\textwidth]{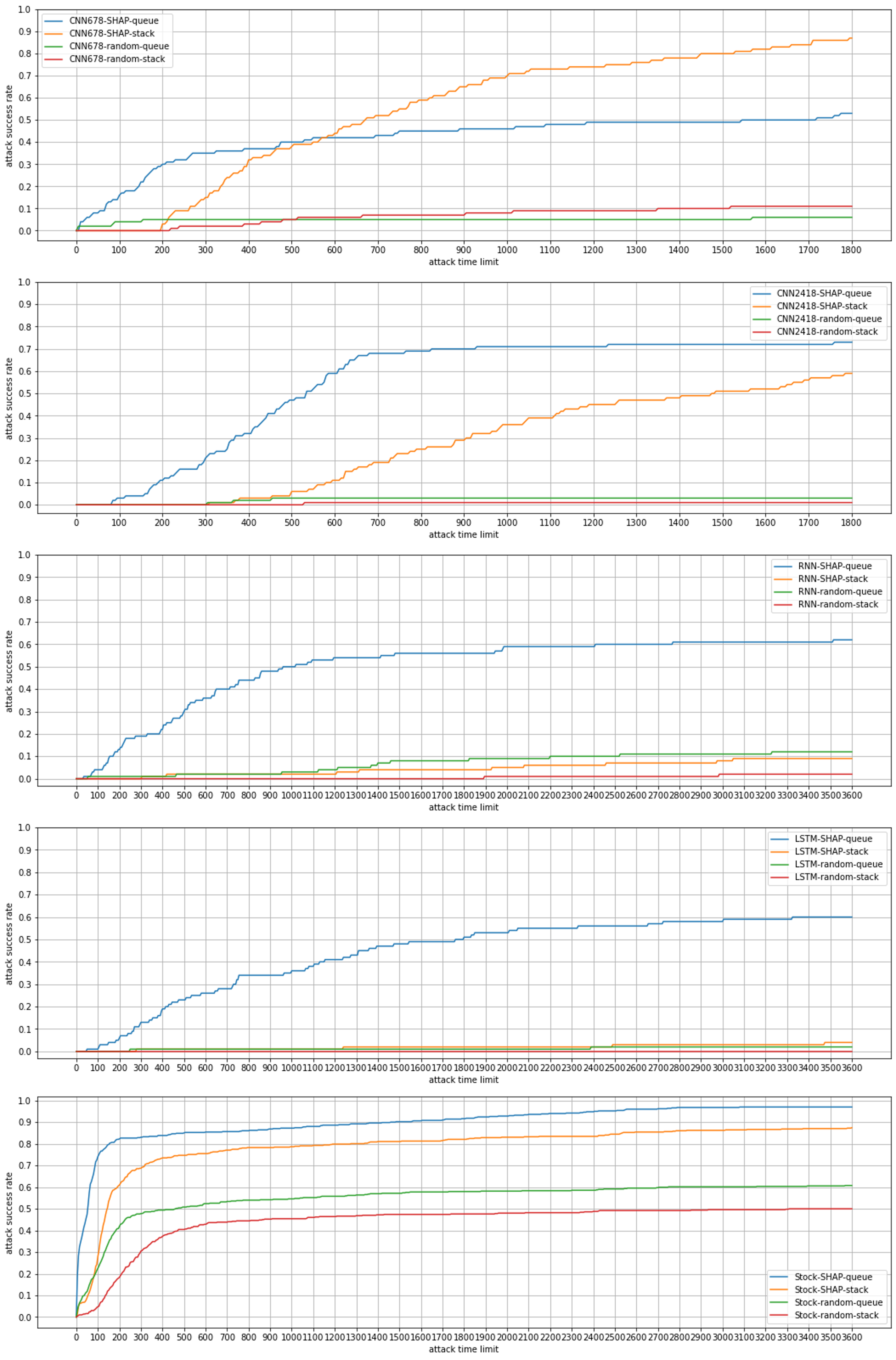}}
\caption{Attack rate vs. Solving time.}
\label{attack_rate}
\end{figure}

\begin{table}
\centering
\caption{Forwarding Performance of IMDb Model(Successful forward / Solve All Cstr / Failed Forward)}
\begin{tabularx}{0.5\textwidth}{lccccc}
\toprule
\textbf{} & \multicolumn{4}{c}{\textbf{IMDb}} \\
\cmidrule(r){2-5} 
\textbf{Pixel} & \textbf{test cases(\#)} &\textbf{SAT} & \textbf{UNSAT} & \textbf{gen cstr(F time)}\\
\midrule
1 & 5 / 2 / 0 & 24 / 32 & 9225 / 11686 & 1162(947) / 355(783) \\
10 & 6 / 0 / 1 & 2 / NA & 106 / NA & 4733(20798) / NA \\
100 & 0 / 0 / 7 & NA & NA & NA (over 12hrs) \\
1000 & 0 / 0 / 7 & NA & NA & NA (over 12hrs) \\
\bottomrule
\end{tabularx}
\label{tab0}
\end{table}
\subsection{Comparison between Stack and Queue(RQ2)}
In this session, we aim to compare the effectiveness of attacks using queue or stack-based constraint-solving orders. Therefore, we employed SHAP values to select target pixels for attacks across all six models. The statistical results, presented in Table~\ref{tab2}, include the success rate of attacked test cases (ATK) and the average time taken for successful attacks (time). As the attack attempts on the IMDb model were unsuccessful, the outcomes pertaining to IMDb have been excluded from the statistical analysis.

In Fig.~\ref{attack_rate}, we depict the trend charts for the success rates of attacked test cases over time, ranging from 0 to the timeout period. This Figure involves five models, employing queue/stack-based constraint-solving, and utilizing random/SHAP values for selecting target pixels, resulting in four distinct combinations. The results indicate that, whether employing SHAP values or randomly selecting target pixels, a queue-based constraint-solving order yields superior performance in RNN/LSTM models. This is evident from the considerably higher success rates observed with green lines (random \& Queue) compared to red lines (random \& Stack), as well as the analogous trend between blue lines (SHAP \& Queue) and orange lines (SHAP \& Stack). 

In scenarios with sufficient time, we observed distinctly different outcomes between CNN models and RNN/LSTM models. In the results for CNN models, we noted that utilizing a queue-based constraint-solving order led to a rapid initial increase in the success rate of attacks, followed by a quick stabilization. Particularly in the case of the CNN\_678 model, around time = 600 seconds, the attack performance of the stack method started to surpass that of the queue. Furthermore, in the results for the CNN\_2418 model, we observed that the attack success rate for the stack method continued to grow rapidly until reaching the timeout, while the attack success rate for the queue method had already leveled off. We hypothesize that the chosen timeout duration might not be sufficiently long to demonstrate the point at which the stack method outperforms the queue.

We can distinctly derive two conclusions from Table~\ref{tab2} and Fig.~\ref{attack_rate}. First, when employing the CNN model, the performance of the stack-based attack surpasses that of the queue. Conversely, in the case of RNN or LSTM models, the queue outperforms the stack. The second conclusion indicates that, irrespective of the model used, the average duration for successful attacks with the queue is shorter than that with the stack. This is attributed to the queue's tendency to generate constraints for the frontmost layers first, resulting in shorter and fewer constraints per set, contributing to this observed outcome.

\textbf{Answer to RQ2:} Drawing upon the presented statistics, we arrive at the following conclusions: for RNN/LSTM models, a strongly recommended approach involves adopting a queue-based constraint-solving order. In the case of CNN models, if prioritizing a high success rate is paramount, a stack-based constraint-solving order is advisable. However, if minimizing the attack time takes precedence over success rate, opting for a queue-based constraint-solving order proves to be a more effective choice.

\begin{table*}
\centering
\caption{Attack result on Queue and Stack(SHAP)}
\begin{tabularx}{0.82\textwidth}{lccccccccccccc}
\toprule
\textbf{} & \multicolumn{2}{c}{\textbf{CNN$\_$678}} & \multicolumn{2}{c}{\textbf{CNN$\_$2418}} & \multicolumn{2}{c}{\textbf{RNN}} & \multicolumn{2}{c}{\textbf{LSTM}} & \multicolumn{2}{c}{\textbf{Stock}} \\
\cmidrule(r){2-3} \cmidrule(r){4-5} \cmidrule(r){6-7} \cmidrule(r){8-9} \cmidrule(r){10-11} 
\textbf{Order} & \textbf{ATK(\%)} & \textbf{time} & \textbf{ATK(\%)} & \textbf{time} & \textbf{ATK(\%)} & \textbf{time} & \textbf{ATK(\%)} & \textbf{time} & \textbf{ATK(\%)} & \textbf{time} \\
\midrule
Queue & 53\% & 386.28 & 73\% & 521.48 & 62\% & 702.45 & 60\% & 696.52 & 97.01\% & 254.88 \\
Stack & 87\% & 688.81 & 59\% & 960.39 & 9\% & 1747.73 & 4\% & 1869.22 & 87.45\% & 357.37 \\
\bottomrule
\end{tabularx}
\label{tab2}
\end{table*}

\subsection{Influence using SHAP value(RQ3)}
In this section, our objective is to assess the efficacy of attacks employing random or SHAP values in selecting target pixels. To achieve this, we utilized a queue-based constraint-solving approach for attacks conducted on all six models. The statistical outcomes, outlined in Table~\ref{tab3}, encompass the success rate of attacked test cases (ATK) and the average time required for successful attacks (time). Since there were no successful attack instances on the IMDb model, the statistical data does not incorporate results related to IMDb. 

In Fig.~\ref{attack_rate}, the significantly elevated success rates depicted by the blue lines (SHAP \& Queue) in contrast to the green lines (random \& Queue), as well as the similar pattern observed between the orange lines (SHAP \& Stack) and red lines (random \& Stack), are apparent.

\textbf{Answer to RQ3:} In terms of attack success rates, employing SHAP values for selecting target pixels consistently yields superior performance. As for the average time required for successful attacks, randomly selecting target pixels occasionally matches the time taken when using SHAP values. However, more often than not, the use of SHAP values significantly reduces the time required for attacks.

\begin{table*}
\centering
\caption{Attack result on Random and SHAP(Queue)}
\begin{tabularx}{0.82\textwidth}{lccccccccccccc}
\toprule
\textbf{} & \multicolumn{2}{c}{\textbf{CNN$\_$678}} & \multicolumn{2}{c}{\textbf{CNN$\_$2418}} & \multicolumn{2}{c}{\textbf{RNN}} & \multicolumn{2}{c}{\textbf{LSTM}} & \multicolumn{2}{c}{\textbf{Stock}} \\
\cmidrule(r){2-3} \cmidrule(r){4-5} \cmidrule(r){6-7} \cmidrule(r){8-9} \cmidrule(r){10-11} 
\textbf{Pick} & \textbf{ATK(\%)} & \textbf{time} & \textbf{ATK(\%)} & \textbf{time} & \textbf{ATK(\%)} & \textbf{time} & \textbf{ATK(\%)} & \textbf{time} & \textbf{ATK(\%)} & \textbf{time} \\
\midrule
Random & 6\% & 317.29 & 3\% & 371.9 & 12 \% & 1483.05 & 2\% & 1318.13 & 60.76\% & 339.43 \\
SHAP & 53\% & 386.29 & 73\% & 521.48 & 62\% & 702.45 & 60\% & 696.52 & 97.21\% & 139.46 \\
\bottomrule
\end{tabularx}
\label{tab3}
\end{table*}

\subsection{Constraints-solving performance(RQ4)}
In this session, we aim to investigate the impact of PyCT across different models and varying quantities of target pixels on constraint resolution performance. To achieve this, we compiled attack data for all six models, considering both random and SHAP values for selecting target pixels, as well as queue/stack-based constraint-solving orders. The statistical results, presented in Table~\ref{tab4}, include the average time required to resolve each set of constraints (time) and the average file size for each set of constraints (size).

From Table~\ref{tab4}, we observe that, across the attack statistics for each model, the file size of constraints and the time spent on constraint resolution consistently increase with the growing number of target pixels. Furthermore, in the statistical analysis of two CNN models with identical structures but differing parameters, we note that, for the same number of target pixels, models with more parameters (i.e., greater complexity) exhibit larger constraint file sizes and require more time for constraint resolution.

\textbf{Answer to RQ4:} Based on the above statistical data, we draw the following conclusions: the file size of constraints increases with the growing number of target pixels and also escalates with the rising complexity of the model. Consequently, this results in an increased time expenditure for PyCT in constraint resolution. However, in our experiments, the majority of time expenditures were quite brief, with the average duration rarely exceeding 0.5 seconds.

\begin{table*}
\centering
\caption{Constraint solving efficiency}
\begin{tabularx}{1\textwidth}{lccccccccccccc}
\toprule
\textbf{} & \multicolumn{2}{c}{\textbf{CNN$\_$678}} & \multicolumn{2}{c}{\textbf{CNN$\_$2418}} & \multicolumn{2}{c}{\textbf{RNN}} & \multicolumn{2}{c}{\textbf{LSTM}} & \multicolumn{3}{c}{\textbf{Stock}} & \multicolumn{2}{c}{\textbf{IMDb}} \\
\cmidrule(r){2-3} \cmidrule(r){4-5} \cmidrule(r){6-7} \cmidrule(r){8-9} \cmidrule(r){11-12} \cmidrule(r){13-14}
\textbf{Pixel} & \textbf{time(s)} & \textbf{size(KB)}& \textbf{time(s)} & \textbf{size(KB)} & \textbf{time(s)} & \textbf{size(KB)} & \textbf{time(s)} & \textbf{size(KB)} & \textbf{Pixel} & \textbf{time(s)} & \textbf{size(KB)} & \textbf{time(s)} & \textbf{size(KB)} \\
\midrule
1 & 0.09 & 71.8 & 0.2 & 225.02 & 0.06 & 11.19 & 0.11 & 58.33 & 1 & 0.07 & 23.77 & 3.4 & 2134 \\
4 & 0.13 & 117.09 & 0.3 & 339.13 & 0.1 & 36.86 & 0.28 & 201.12 & 2 & 0.09 & 51.28 & 3.5 & 2256 \\
8 & 0.17 & 156.62 & 0.39 & 429.86 & 0.14 & 67.54 & 0.43 & 332.66 & 3 & 0.12 & 73.16 & 3.4 & 2072 \\
16 & 0.23 & 197.71 & 0.55 & 546.35 & 0.21 & 116.55 & 0.69 & 547.43 & 4 & 0.13 & 91.65 & 3.7 & 2284 \\
32 & 0.68 & 249.16 & 1.16 & 681.67 & 0.32 & 194.49 & 0.96 & 769.15 & 8 & 0.19 & 150.57 & 4.2 & 2458 \\
\bottomrule
\end{tabularx}
\label{tab4}
\end{table*}

\subsection{Comparison with DeepConcolic(RQ5)}
As the exclusive concolic testing tool designed specifically for neural networks up to the present, we chose DeepConcolic~\cite{b14} for comparison with PyCT in our experimental design. Recognized for its specialized focus on attacks exclusively targeting CNN models, DeepConcolic emerges as a distinctive candidate in our study. In the DeepConcolic attack process, a random image is selected as a seed input (referred to as the initial input in PyCT) from the Mnist test dataset comprising 10,000 images. This input is then fed into a specific neural network model for a forward pass. DeepConcolic identifies the neuron most likely to be activated in the activation layer, considering input values $\geq$ 0 as activated (ReLU layer criterion) and $<$ 0 as non-activated. Among negative values, the one closest to 0 is deemed by DeepConcolic as the most likely to be activated. This neuron is then designated as the target neuron, and a constraint is generated to affect this neuron. The constraint is subsequently fed into a solver to produce a new test input that has the potential to activate the target neuron, thus enhancing neuron coverage.

In our experiments, as DeepConcolic is designed to randomly select seed inputs, we initiated attacks on two pre-trained Mnist CNN models (CNN\_678 and CNN\_2418) using DeepConcolic with 1000 images, ensuring no repetitions. Subsequently, the same set of 1000 images was subjected to attacks with PyCT. To ensure comparable attack times of 10 minutes for both approaches, we conducted 100 trial runs with DeepConcolic to determine the maximum iterations that would result in each run exceeding 10 minutes. Following this, for the statistical analysis, we selected results within a 10-minute timeframe for both PyCT and DeepConcolic, ensuring equal attack times. The GitHub repository for DeepConcolic can be found at https://github.com/TrustAI/DeepConcolic, and the command used for DeepConcolic is as follows: 'python -m deepconcolic.main --model \{model\_name\}.h5 --outputs out/\{model\_name\} --max-iterations \{max\} --dataset mnist --criterion nc --norm linf --save-all-tests'. Here, \{model\_name\} is replaced with the name of the model used in the attack, and \{max\} is filled with the determined maximum iteration quantity that ensures a runtime exceeding 10 minutes.

In the context of PyCT attacks, we commenced by selecting a single target pixel based on SHAP values, with a timeout set at 10 minutes for each attack attempt. Following unsuccessful test cases from the initial attack, four target pixels were chosen, and this process was iteratively repeated, increasing the number of target pixels in increments of 8, 16, and ultimately up to 32. Attacks were conducted for each set of chosen pixels until the completion of the entire sequence.

After attacking 1000 images from the MNIST dataset using DeepConcolic and PyCT on CNN\_678 and CNN\_2418 models, the comparison of the attack results is presented in Table~\ref{tab6}, with a timeout of ten minutes for each attack. The experimental description encompasses the cumulative success rate of PyCT, the success rate of DeepConcolic (ATK), and the average time required for a successful attack (time). While DeepConcolic requires significantly less time for a successful attack, it is noteworthy that PyCT demonstrates a higher attack rate when utilizing SHAP values. Notably, PyCT exhibits improved attack performance even when selecting only a small amount of target pixels. However, it is important to highlight that, despite the higher success rate with SHAP values, PyCT faced challenges in certain test cases, failing to achieve successful attacks all due to timeouts. Additionally, none of the test cases solved all of the constraint conditions, indicating that PyCT may face constraints related to computational resource limitations in specific scenarios, preventing the completion of attacks within the specified time frame. This observation suggests that with more powerful computational resources, PyCT's attack performance could potentially be further enhanced. 

\textbf{Answer to RQ5:} Based on the above statistics, we can conclude that when using the current model, PyCT achieves significantly better attack performance by modifying only a small amount of pixels compared to DeepConcolic, which has the capability to modify all pixels. However, in terms of attack success time, DeepConcolic demonstrates remarkable efficiency, achieving successful attacks in a minimal amount of time for each image. As a recommendation, when targeting the inputs of a CNN model for attacks, it may be advisable to initially use DeepConcolic for a brief period to attack a large number of inputs. Subsequently, for unsuccessful test inputs, PyCT can be employed gradually, increasing the number of target pixels to enhance the attack success rate.

\begin{table}
\centering
\caption{Attack performance on PyCT and DeepConcolic}
\begin{tabularx}{0.47\textwidth}{lccccccccccccc}
\toprule
\textbf{} & \textbf{} & \multicolumn{2}{c}{\textbf{CNN$\_$678}} & \multicolumn{2}{c}{\textbf{CNN$\_$2418}} \\
\cmidrule(r){3-4} \cmidrule(r){5-6}
\textbf{Tool} & \textbf{Pixel} & \textbf{ATK(\%)} & \textbf{time (s)} & \textbf{ATK(\%)} & \textbf{time (s)}\\
\midrule
 & 1 & 1.8\% & 162.5 & 0.8\% & 378.49 \\
 & 4 & 3.5\% & 234.22 & 1.4\% & 372.66 \\
PyCT & 8 & 15.8\% & 259.56 & 5.7\% & 394.96 \\
 & 16 & 44.7\% & 281.52 & 17.4\% & 466.78 \\
 & 32 & 71.1\% & 316.82 & 18\% & 526.31 \\
\midrule
DeepConcolic & all & 7.2\% & 4.94 & 1.1\% & 30.36 \\
\bottomrule
\end{tabularx}
\label{tab6}
\end{table}

\section{Conclusion}
In this research, we propose constraint based adversarial example synthesis and realize the idea with PyCT, a Python concolic testing tool, to encompass various neural network operations. The contribution includes achieving comprehensive support for key operations like floating-point computation, connected layer computation, convolution, recurrent neural networks, and long short-term memory networks.

These enhancements empower PyCT to systematically explore pivotal branch decisions influencing network predictions and generate crucial test inputs for adversarial example creation. Through our efforts, we spotlighted vulnerabilities in neural networks along critical decision paths, particularly concerning non-convex activation functions like ReLU, tanh, and Sigmoid. This holistic approach evaluates neural network security, underscoring the importance of rigorous testing and fortification.

Looking ahead, future work could focus on optimizing PyCT to reduce computational demands, enabling it to achieve impactful results in shorter timeframes. This approach would not only enhance PyCT's efficiency but also address the computational challenges associated with testing increasingly complex neural network structures.

\end{document}